\documentclass[useAMS,usenatbib]{mnras}
\usepackage{graphicx}
\usepackage{amsmath}
\usepackage{amssymb}
\usepackage{color}

\newcommand \etal {et~al.~}

\def \spose#1{\hbox  to 0pt{#1\hss}}  
\newcommand \lta{\mathrel{\spose{\lower 3pt\hbox{$\sim$}}\raise  2.0pt\hbox{$<$}}}
\newcommand \gta{\mathrel{\spose{\lower  3pt\hbox{$\sim$}}\raise 2.0pt\hbox{$>$}}}

\newcommand{\Lsun}{{\ifmmode{ {\rm L}_{\odot}}\else{L_{\odot}}\fi}}
\newcommand{\Msun}{{\ifmmode{ {\rm M}_{\odot}}\else{M_{\odot}}\fi}}
\newcommand{\Mstar}{{\ifmmode{M_{\rm star}}\else{$M_{\rm star}$}\fi}}
\newcommand{\kms}{{\ifmmode{ {\rm km\,s^{-1}} }\else{ ${\rm km\,s^{-1}}$ }\fi}}
\newcommand{\fbar}{\ifmmode{f_{\rm bar}}\else{$f_{\rm bar}$}\fi}
\newcommand \Omegadm {\ifmmode \Omega_{\rm dm} \else $\Omega_{\rm dm}$ \fi}
\newcommand \Omegam {\ifmmode \Omega_{\rm m} \else $\Omega_{\rm m}$ \fi} 
\newcommand \Omegab {\ifmmode \Omega_{\rm b} \else $\Omega_{\rm b}$ \fi} 
\newcommand \OmegaL {\ifmmode \Omega_{\rm \Lambda} \else $\Omega_{\rm \Lambda}$\fi} 
\newcommand \Deltavir {\ifmmode \Delta_{\rm vir} \else $\Delta_{\rm vir}$ \fi}
\newcommand \rhocrit {\ifmmode \rho_{\rm crit} \else $\rho_{\rm crit}$ \fi}

\title[Star formation threshold and halo response]{NIHAO XX: The impact of the star formation threshold on the cusp-core transformation of cold dark matter haloes}
\author[Dutton \etal]{Aaron A. Dutton$^1$\thanks{dutton@nyu.edu},
  Andrea V. Macci\`{o}$^{1,2}$, Tobias Buck$^2$, Keri L. Dixon$^1$, 
  \newauthor{Marvin Blank$^{1,3}$, Aura Obreja$^{4}$}\\
 $^1$New York University Abu Dhabi, PO Box 129188, Saadiyat Island, Abu Dhabi, United Arab Emirates\\
 $^2$Max Planck Institut f\"{u}r Astronomie, K\"{o}nigstuhl 17, 69117 Heidelberg, Germany\\
 $^3$Institut f\"{u}r Theoretische Physik und Astrophysik, Christian-Albrechts-Universit\"{a}t zu Kiel, Leibnizstr. 15, D-24118 Kiel, Germany\\
 $^4$University Observatory Munich, Scheinerstra\ss e 1, D-81679 Munich, Germany\\
}

\begin{document}

\maketitle

\label{firstpage}

\begin{abstract}
We use cosmological hydrodynamical galaxy formation simulations from
the NIHAO project to investigate the impact of the threshold for star
formation on the response of the dark matter (DM) halo to baryonic
processes.  The fiducial NIHAO threshold, $n=10\,[{\rm cm}^{-3}]$,
results in strong expansion of the DM halo in galaxies with stellar
masses in the range $10^{7.5} \lta \Mstar \lta 10^{9.5} \Msun$.  We
find that lower thresholds such as $n=0.1$ (as employed by the
EAGLE/APOSTLE and Illustris/AURIGA projects) do not result in
significant halo expansion at any mass scale.  Halo expansion driven
by supernova feedback requires significant fluctuations in the local
gas fraction on sub-dynamical times (i.e., $\lta$ 50 Myr at galaxy
half-light radii), which are themselves caused by variability in the
star formation rate.  At one per cent of the virial radius,
simulations with $n=10$ have gas fractions of $\simeq 0.2$ and
variations of $\simeq 0.1$, while $n=0.1$ simulations have order of
magnitude lower gas fractions and hence do not expand the halo.  The
observed DM circular velocities of nearby dwarf galaxies are
inconsistent with CDM simulations with $n=0.1$ and $n=1$, but in
reasonable agreement with $n=10$. Star formation rates are more
variable for higher $n$, lower galaxy masses, and when star formation
is measured on shorter time scales.  For example, simulations with
$n=10$ have up to 0.4 dex higher scatter in specific star formation
rates than simulations with $n=0.1$.  Thus observationally
constraining the sub-grid model for star formation, and hence the
nature of DM, should be possible in the near future.
\end{abstract}


%% SECTION 1
\section{Introduction}

The structure of dark matter haloes on kiloparsec-scales potentially
provides the most sensitive astrophysical test of the cold dark matter
(CDM) paradigm, and more generally the nature of dark matter
\citep[e.g.,][]{Bullock17}.  In dissipationless CDM simulations, the
halo structure is well determined \citep[e.g.,][]{Stadel09,Dutton14}.
Gas dissipation is thought to only make the dark matter halo contract
\citep{Blumenthal86,Gnedin04}. However, other baryonic processes can
cause the dark matter halo to expand
\citep[e.g.,][]{El-Zant01,Weinberg02,Read05,Pontzen12}. Due to the
nonlinear nature of these processes and the importance of realistic
initial conditions and evolution, their impact is best studied using
cosmological hydrodynamical simulations.

%% Table 1
\begin{table*}
\begin{center}
  \caption{Simulation parameters. Box is the size of the parent DMO
    simulation. The halo ID corresponds to the virial mass of the
    parent DMO simulation. Number gives the number of haloes we run at
    a given resolution level, which is specified by the dark matter
    and gas particle masses, $m$, and gravitational force softenings,
    $\epsilon$.}
\begin{tabular}{ccccccc}
\hline
Box          & Halo ID range       & Number &$m_{\rm gas}$ & $\epsilon_{\rm gas}$ & $m_{\rm DM}$ & $\epsilon_{\rm DM}$ \\
($h^{-1}$Mpc) & ($h^{-1}$ M$_\odot$) &        &(M$_\odot$)   & (pc)               &   (M$_\odot$)& (pc) \\
\hline 
  60 & g7.08e11 - g8.26e11 & 4 & 3.166$\times$10$^5$  & 397.9 & 1.735$\times$10$^6$  & 931.4 \\
  60 & g1.37e11 - g3.71e11 & 8 & 3.958$\times$10$^4$  & 199.0 & 2.169$\times$10$^5$  & 465.7 \\
  60 & g2.34e10 - g4.99e10 & 4 & 1.173$\times$10$^4$  & 132.6 & 6.426$\times$10$^4$  & 310.5 \\
  20 & g7.05e09 - g1.23e10 & 4 & 3.475$\times$10$^3$  & $\phantom{1}$88.4 & 1.904$\times$10$^4$  & 207.0 \\
 \hline
\end{tabular}
\label{tab:param}
\end{center}
\end{table*}

Early simulations suffered from overcooling, forming an order of
magnitude too many stars. The current generation of simulations is
able to form galaxies with realistic amounts of stars and cold gas
both today and in the past
\citep{Hopkins14,Marinacci14,Wang15,Schaye15}.  Improved numerical
resolution has certainly helped, but the main difference is due to
improved sub-grid models for star formation and feedback.  Sub-grid
models are necessary due to both the large dynamic range required to
simulate cosmological scales and the formation of individual stars and
the fact that the physics of star formation is still an unresolved
problem.  The sub-grid model attempts to capture the process by which
gas turns into stars and the subsequent feedback of energy into the
inter stellar medium (ISM) using {\it effective} models (i.e.,
averaged over many star formation and feedback events).  Note that we
do not necessarily need to understand how stars form on a micro level,
but we do need an effective model of how stars form on kpc scales.
Effective models are common in physics, for example, one can model
atomic processes without an understanding of quarks.

Sub-grid models for star formation  and feedback have several free
parameters, which must currently be calibrated against observations
\citep[e.g.,][]{Schaye15}. In this paper we focus on a single
parameter, the threshold for star formation, $n$. We choose this,
because it is common among most sub-grid galaxy formation models,
because it varies greatly $0.01\lta n\lta 100\,[{\rm cm}^{-3}]$ in
simulations with comparable numerical resolution, and because we have
good reason to suspect it will influence how the dark halo responds to
star formation \citep[e.g.,][]{Pontzen12}.  Furthermore, halo
expansion is only seen in simulations that adopt a high threshold
$n\gta 10$ \citep{Governato10, Maccio12, Pontzen12,
  Teyssier13, DiCintio14, Chan15, Read16, Tollet16}, while simulations
with a low threshold $n\lta 0.1$ never find significant
halo expansion \citep{Oman15, Schaller15}. 

The goal of this paper is to test how the halo response depends on the
threshold for star formation, and to determine any observational ways
to distinguish between different thresholds. This paper is organized
as follows. The simulation suite is outlined in section 2. Results for the
stellar to halo masses and dark matter density profiles are shown in
section 3. Section 4 gives a comparison between simulated and observed dark
matter circular velocity profiles. Section 5 discusses the physical
mechanism that drives the differences in halo response in our
simulations and observational tests. A summary is given in section 6.

%% SECTION 2
\section{Simulations}
\label{sec:sims}

We use a set of 20 haloes of virial masses between  $\sim 10^{10}$ to
$\sim 10^{12} \Msun$ drawn from the NIHAO project \citep{Wang15}, and
re-simulate them with three different $n$.  Here we give a brief overview
of the NIHAO simulations and we refer the reader to \citet{Wang15} for a
more complete discussion.

NIHAO  is a sample of $\sim$ 100 hydrodynamical cosmological zoom-in
simulations using the SPH code {\sc gasoline2}
\citep{Wadsley17}. Haloes are selected at redshift $z=0.0$ from parent
dissipationless simulations of box size 60, 20, and 15 $h^{-1}$Mpc,
presented in \citet{Dutton14}, which adopt a flat $\Lambda$CDM
cosmology with parameters from the \citet{Planck14}: Hubble parameter
$H_0$= 67.1 \kms Mpc$^{-1}$; matter density $\Omegam=0.3175$; dark
energy density $\OmegaL=1-\Omegam=0.6825$; baryon density
$\Omegab=0.0490$; power spectrum normalization $\sigma_8 = 0.8344$,
and power spectrum slope $n=0.9624$.  The corresponding cosmic baryon
fraction $\fbar\equiv \Omegab/\Omegam = 0.154$.  Haloes are selected
uniformly in log halo mass from $\sim 10$ to $\sim 12$ {\it without}
reference to the halo merger history, concentration or spin parameter.

\subsection{Resolution} 
Dark matter particle masses and force softenings are chosen to resolve
the mass profile at $\lta 1$ per cent of the virial radius according
to the \citet{Power03} criteria. This choice results in the dark
matter profile being converged at $\simeq 2$ softening lengths and
$\sim 10^6$ dark matter particles inside the virial radius of all main
haloes at $z = 0$.  The corresponding gas particle masses and force
softenings are a factor of $\Omegab/\Omegadm=0.182$ and
$\sqrt{\Omegab/\Omegadm}=0.427$ lower. The particle masses and force
softenings are given in Table.~\ref{tab:param}.  Each hydro simulation
has a corresponding dark matter only (DMO) simulation of the same
resolution. These simulations have been started using the identical
initial conditions, replacing baryonic particles with dark matter
particles.

The simulations employ adaptive time steps. We start with 1024 major
time steps each of  13.5 million years (Age of universe/1024), then
the code refines this time step according to the acceleration of a
particle.  We allow for a maximum of 20 refinements which sets the
minimum to 12.9 years (maximum time step/$2^{20}$).
   
\subsection{Star formation}
Star formation is implemented as described in
\citet{Stinson06, Stinson13}. 
Stars form from cool ($T < 15 000$K), dense gas ($\rho > n$[cm$^{-3}$]).
Gas eligible to form stars is converted into stars according to 
\begin{equation}
  \frac{\Delta\Mstar}{\Delta t} = c_{\ast} \frac{M_{\rm gas}}{t_{\rm dyn}}.
\end{equation}
Here $\Delta\Mstar$ is the mass of stars formed, $\Delta t=0.84\,$Myr
is the time-step between star formation events, and $t_{\rm dyn}$ is
the gas particle's dynamical time. The efficiency of star formation is
set to $c_{\ast}=0.10$ for all simulations.

The main parameter of relevance to our study is the star formation
threshold, $n$. In our fiducial NIHAO simulations we adopt $n=10 [{\rm
    cm}^{-3}] \simeq 50 m_{\rm gas}/\epsilon_{\rm gas}^3$.  Here 50 is
the number of particles used in the SPH smoothing kernel, $m_{\rm
  gas}$ is the initial mass of gas particles, and $\epsilon_{\rm gas}$
is the gravitational force softening of the gas particle.  In our
simulations we choose $\epsilon_{\rm gas}\propto m_{\rm gas}^{1/3}$,
so that the star formation threshold is {\it independent} of the gas
particle mass.  We run each simulation at two additional star
formation thresholds: $n=0.1$ and $n=1.0$.  The former is similar to
that adopted by the EAGLE/APOSTLE \citep{Schaye15, Sawala16} and
Illustris/AURIGA \citep{Vogelsberger14, Grand17} projects.  The FIRE
project \citep{Hopkins14,Hopkins18} uses star formation thresholds as
high as $n=100$ or $n=1000$. We do not run simulations with such high
values because our choice of gas mass and force softening do not
enable us to resolve these densities.  All simulations in the NIHAO
project including the ones used here employ a pressure floor to keep
the Jeans mass of the gas resolved and suppress artificial
fragmentation.

Note that all of the thresholds we try are well below the density
where stars form in the real Universe. Stars form in the cores of
giant molecular clouds at densities of $\rho\sim 10^4\,[{\rm
  cm}^{-3}]$. Giant molecular clouds have typical densities of $\rho\sim
100 \,[{\rm cm}^{-3}]$, while molecular gas typically forms at densities
$\rho\gta 10 \,[{\rm cm}^{-3}]$.  Thus it might be surprising if star
formation could be modeled accurately on kpc scales using a density
threshold as low as $n=0.1$. However, as discussed in
the introduction, we are not modeling individual star formation and SN
events, but rather an ensemble of them. So it might be possible to
capture the essential features of star formation and feedback using an
effective star formation threshold that is much lower than that for
individual star forming regions.

\subsection{Feedback}
The NIHAO simulations employ thermal feedback in two epochs following
\citet{Stinson13}.  In the first epoch, ‘pre-SN feedback’ (early
stellar feedback, ESF) happens before any supernovae explode. The ESF
represents stellar winds and photoionization from the bright young
stars. The ESF consists of a fraction $\epsilon_{\rm ESF}$ of the
total stellar flux being ejected from stars into surrounding gas ($2
\times 10^{50}$ erg of thermal energy per $\Msun$ of the entire
stellar population). Radiative cooling is left on for the pre-SN
feedback.  The second epoch starts 4 Myr after the star forms, when
the first supernovae start exploding. Only supernova energy is
considered as feedback in this second epoch. Stars with mass $8\,\Msun
< M_{\ast} < 40\,\Msun$ eject both energy ($\epsilon_{\rm SN}\times
10^{51}$ erg/SN) and metals into the interstellar medium gas
surrounding the region where they formed. Supernova feedback is
implemented using the blastwave formalism described in
\citet{Stinson06}. Since the gas receiving the energy is dense, it
would be quickly radiated away due to its efficient cooling. For this
reason, cooling is delayed for particles inside the blast region for
$\sim 30$ Myr.  The free parameters of the feedback model
$\epsilon_{\rm ESF}=0.13, \epsilon_{\rm SN}=1.0$ were calibrated
against the evolution of the stellar mass versus halo mass relation
from halo abundance matching \citep{Behroozi13,Moster13} for a $z=0$
Milky Way mass halo $\sim 10^{12}\Msun$.
  
In EAGLE, energy feedback from star formation is implemented using the
stochastic thermal prescription of \citet{DallaVecchia12}. Rather than
delay cooling, they delay the injection of energy into gas particles
so that those particles that get heated are too hot to cool
efficiently.  The energy injected per unit stellar mass formed
decreases with the metallicity of the gas and increases with  the gas
density to account for unresolved radiative losses and to  help
prevent spurious numerical losses.  The free parameters of the EAGLE
star formation and feedback model are calibrated against the observed,
$z = 0$ galaxy stellar mass function and the size versus stellar mass
relation of $z\simeq 0$ star-forming galaxies \citep{Crain15}.

In NIHAO and EAGLE galactic winds develop naturally, without imposing
mass loading factors, velocities or directions. On the other hand, the
AURIGA simulations adopt a kinetic feedback model in which gas
particles isotropically surrounding a star particle are given a kick
with velocity proportional to the 1D velocity dispersion of the
surrounding dark matter particles and a mass loading factor of 0.6.
All three models described above are obviously approximations of how
feedback actually occurs, yet they are successful in that they form
galaxies with realistic stellar masses and sizes.  

%% FIGURE 1
%------------------------------------------
\begin{figure*}
  \includegraphics[width=0.85\textwidth]{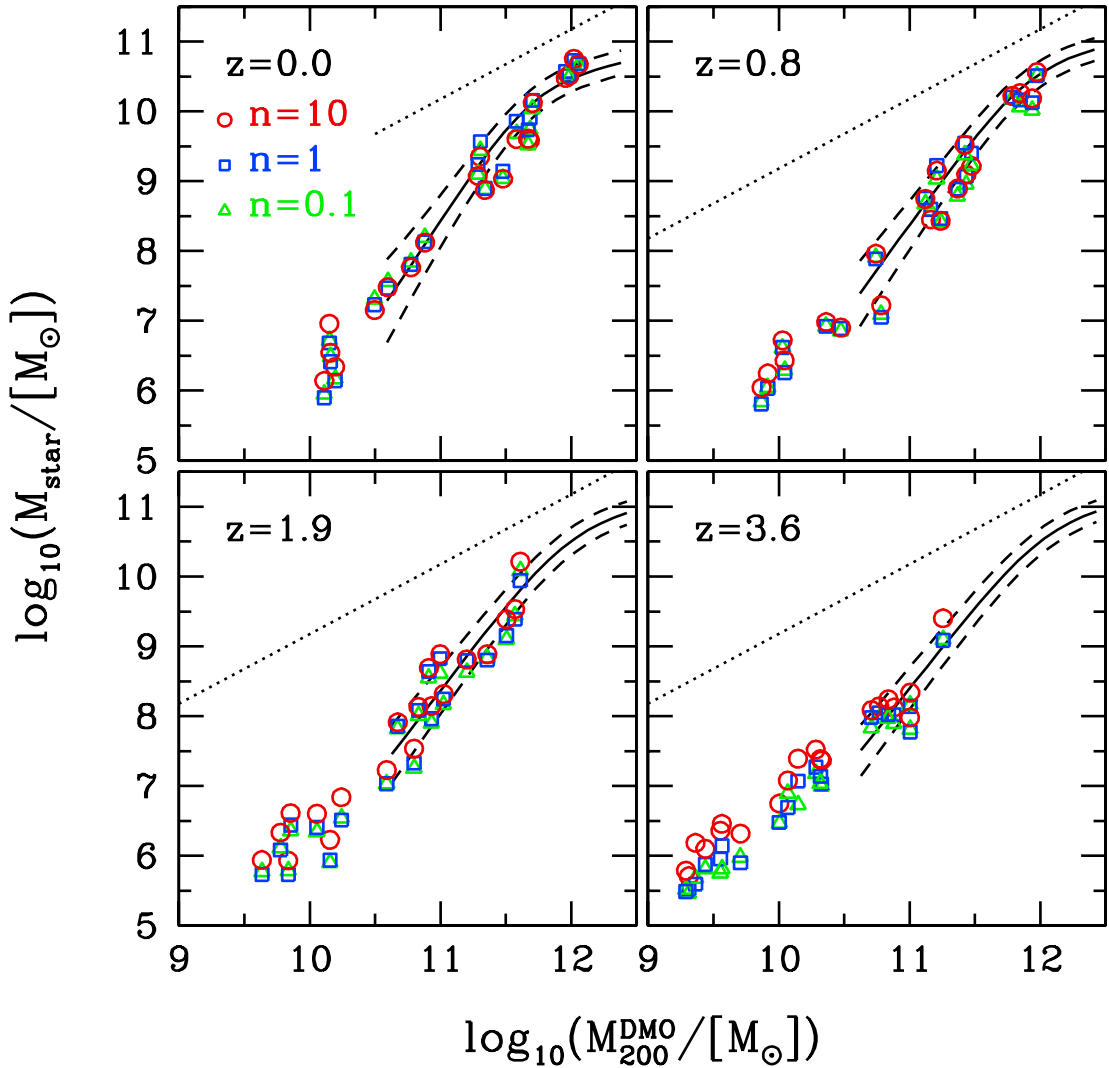}
  \caption{Stellar mass of hydro simulation versus virial mass of DMO
    simulation at four redshifts. Simulations with different star
    formation thresholds ($n$) are shown with different colours and
    symbols: $n=10$ (red circles), $n=1$ (blue squares), $n=0.1$
    (green triangles). We compare to the abundance matching relations
    from \citet{Moster18} (mean and scatter). The dotted line shows the
    cosmic baryon fraction. }
\label{fig:msmh}
\end{figure*}
%------------------------------------------

\subsection{Haloes and galaxies}
Haloes are identified using the MPI+OpenMP hybrid halo finder
\texttt{AHF}\footnote{http://popia.ft.uam.es/AMIGA} \citep{Gill04,
  Knollmann09}. \texttt{AHF} locates local over-densities in an
adaptively smoothed density field as prospective halo centers. The
virial masses of the haloes are defined as the masses within a sphere
whose average density is 200 times the cosmic critical matter density,
$\rhocrit=3H_0^2/8\pi G$.  The virial mass, size and circular velocity
of the hydro simulations are denoted: $M_{200}$, $R_{200}$, $V_{200}$.
The corresponding properties for the DMO simulations are
denoted with a superscript, ${\rm DMO}$.  For the baryons, we calculate
masses enclosed within spheres of radius $r_{\rm gal}=0.2R_{200}$,
which corresponds to $\sim 10$ to $\sim 50$ kpc.  The stellar mass
inside $r_{\rm gal}$ is $M_{\rm star}$, the neutral hydrogen, H{\sc i},
inside $r_{\rm gal}$ is computed following \citet{Rahmati13} as
described in \citet{Gutcke17}.

The NIHAO simulations are the largest set of cosmological zoom-ins
covering the halo mass range $10^{10}$ to $10^{12}\Msun$. Their
uniqueness is in the combination of high spatial resolution coupled to
a statistical sample of haloes.  As discussed in previous papers in
the NIHAO series, NIHAO galaxies are consistent with a wide range of
galaxy properties in both the local and distant Universe.  In the
context of $\Lambda$CDM, they form ``right'' amount of stars both
today and at earlier times \citep{Wang15}. Their cold gas masses and
sizes are consistent with observations \citep{Stinson15, Maccio16,
  Dutton19}, they follow the gas, stellar, and baryonic Tully-Fisher
relations \citep{Dutton17}.  They match the observed clumpy morphology
of galaxies seen at high redshifts \citep{Buck17}.  On the scale of
dwarf galaxies the dark matter haloes expand yielding cored dark
matter density profiles consistent with observations \citep{Tollet16},
and resolve the too-big-to-fail problem of field galaxies
\citep{Dutton16a}.  They reproduce the diversity of dwarf galaxy
rotation curve shapes \citet{Santos-Santos18}, and the H{\sc i}
linewidth velocity function \citep{Maccio16, Dutton19}.  As such they
are a good template with which to predict the structure of cold dark
matter haloes.

%% FIGURE 2
%------------------------------------------
\begin{figure*}
  \includegraphics[width=0.9\textwidth]{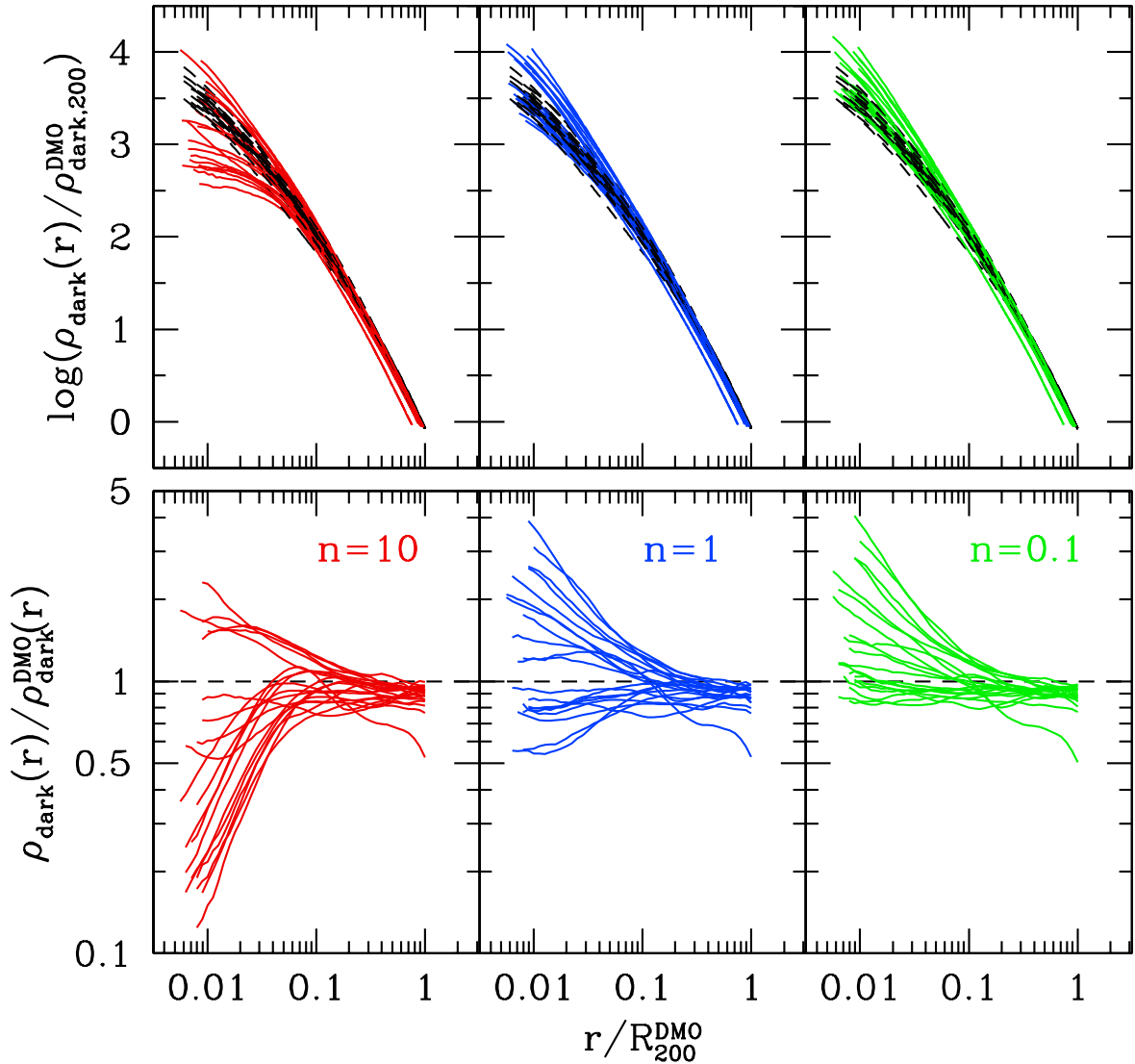}
  \caption{Enclosed dark matter density profiles, where
    $\rho(r)=M(<r)/(4/3 \pi r^3)$.  Dashed lines show DMO
    simulation scaled by $(1-\fbar)=0.85$, solid lines show dark matter
    density profiles from the hydrodynamical simulations with
    $n=10$ (red, left), $n=1$ (blue, middle), and $n=0.1$
    (green, right). Lines are plotted from twice the dark matter
    softening ($\approx$ convergence radius) to the virial radius.}
\label{fig:rho}
\end{figure*}
%------------------------------------------

%% FIGURE 3
%------------------------------------------
\begin{figure*}
  \includegraphics[width=0.9\textwidth]{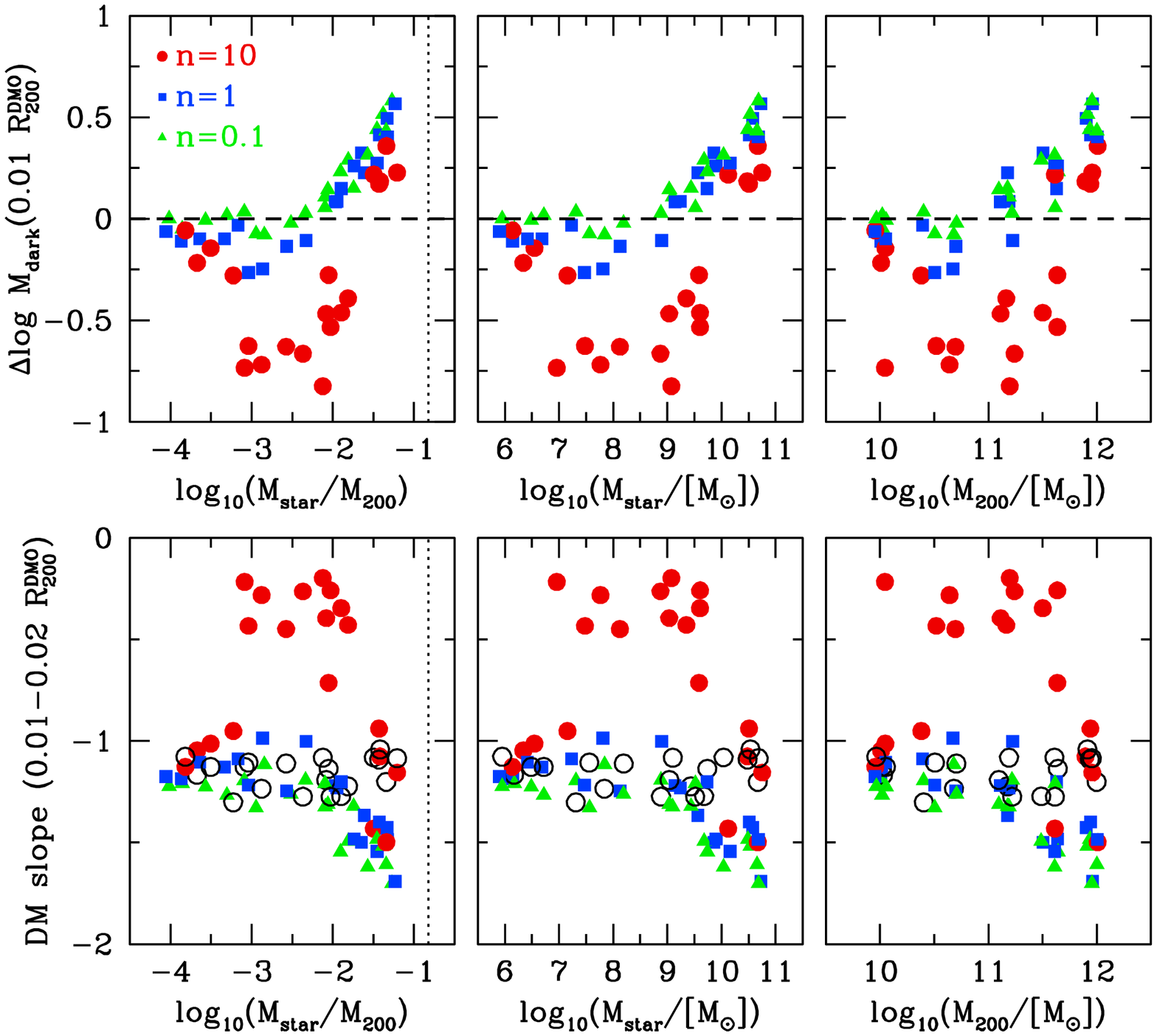}
  \caption{Change in enclosed mass ($\Delta\log M_{\rm dark} \equiv
    \log M_{\rm dark} -\log M_{\rm dark}^{\rm DMO}$) at 1 per cent of
    the virial radius (upper panels) and slope of the enclosed dark
    matter density profile between 1 and 2 per cent of the virial
    radius (lower panels) versus halo mass (right), stellar mass
    (middle), and stellar to halo mass ratio (left). The dotted line
    in the left panels corresponds to the cosmic baryon fraction.}
\label{fig:alpha}
\end{figure*}
%------------------------------------------

%% SECTION 3
\section{Results}

\subsection{Stellar to halo mass}
We start our discussion of the simulations with the two most basic
properties: the total mass within the virial radius of the DMO
simulation $M_{200}^{\rm DMO}$; and the stellar mass (within 20 per
cent of the virial radius) of the corresponding hydrodynamical
simulation, $\Mstar$.  Fig.~\ref{fig:msmh} shows the stellar versus
halo mass relation for the main halo in each of our simulations at
four redshifts: $z=3.6$, $z=1.9$, $z=0.8$, and $z=0.0$.

The different symbols show the three star formation thresholds: $n=10$
(red circles), $n=1$ (blue squares), $n=0.1$ (green triangles). We
maintain this colour and symbol scheme throughout the paper.
Simulations with $n=10$ and $n=1$ have sufficiently similar stellar
masses that we do not re-calibrate the feedback efficiency. However,
for the $n=0.1$ simulations the stellar masses are significantly lower
by redshift $z=0$: a factor of $\sim 2$ for $M_{200}\sim
10^{10}\Msun$, and a factor of $\sim 8$ for $M_{200}\sim
10^{12}\Msun$.  For simplicity we adjust one of the feedback
parameters, $e_{\rm ESF}$, the fraction of the energy from young stars
that couples to the ISM. The fiducial simulations have $e_{\rm
  ESF}=0.13$, which we reduce to $e_{\rm ESF}=0.04$ for the $n=0.1$
simulations, such that the $z=0$ stellar masses are nearly equal to
the corresponding masses in the $n=1$ and $n=10$ cases.  This
re-calibration is an important step that is often not taken.  It
enables us to compare the halo responses of the different thresholds,
without having to worry about other processes.  For example,
\citet{Governato10} simulate a single halo with $n=100$ and
$n=0.1$. The $n=100$ halo expands, while the $n=0.1$ halo
contracts. However, the low threshold simulation formed an order of
magnitude more stars, so it is hard to disentangle the effects of
increased dissipation with the effects of a low threshold.

The dotted lines show the maximum stellar mass, corresponding to all
of the cosmically available baryons turning into stars. The solid and
dashed lines show the mean and $1\sigma$ scatter relations from halo
abundance matching \citep{Moster18}. Here we have converted the halo
masses of the relations to our definition, $M_{200}/(4/3\pi
R_{200}^3)=200 \rhocrit$, using the concentration mass relations from
\citet{Dutton14}.  We have interpolated the fitting coefficients from
\citet{Moster18} to  the redshifts that we show.  Fig.~\ref{fig:msmh}
shows that our simulations with different star formation thresholds
form the ``correct'' (in the context of $\Lambda$CDM) amount of stars
both today and at earlier cosmic times.

\subsection{Density profiles}

Fig.~\ref{fig:rho} shows the enclosed dark matter density profiles,
$\rho(r)\equiv 3 M(<r)/(4\pi r^3)$, for the simulations at $z=0$. We
use enclosed rather than local dark matter density profiles because
the former are more robust being independent of binning.  Coloured
lines correspond to the hydro simulations. Black dashed lines show the
DMO simulations with a rescaling of the total density profile by
$1-\fbar \simeq 0.85$. The re-scaling is done so that when we compare
profiles from hydro and DMO simulations, a ratio of unity corresponds
to no halo response.  Upper panels show the density profiles and lower
panels show the ratio with respect to the DMO simulation. Lines which
are above unity thus correspond to halo contraction, while lines below
unity correspond to halo expansion.  All thresholds result in the same
behaviour at large radii, namely a small amount of (adiabatic) halo
expansion due to the loss of baryons from the halo. The exception is
halo g2.19e11, which has a higher halo mass in DMO due to a major
merger that has been delayed in the hydro simulations.

Below 10 per cent of the virial radius the simulation results start to
diverge, both in the magnitude of the expansion and contraction.  The
$n=0.1$ simulations show a variation of a factor of $\simeq 5$ with
respect to the DMO simulations, the $n=1$ simulations a factor of
$\simeq 6$, and the $n=10$ simulations a factor of $\sim 10$. Thus for
all of the star formation thresholds we try, the cold dark matter
haloes are not described by a universal function, as is the case for
DMO simulations. 

The diversity in halo response at small radii is more clearly shown in
Fig.~\ref{fig:alpha}. Upper panels show the change in the dark matter
mass profile at 1 per cent of the virial radius (identical to the change in
enclosed dark matter density), while lower panels show the slope of
the enclosed dark matter density profile between 1  and 2 per cent of the
virial radius. Note that here we use the enclosed dark matter density
rather than the  local dark matter density as used in our previous
works \citep[e.g.,][]{Tollet16}, but the results are qualitatively the same.

In the upper panels, the dashed line corresponds to the DMO simulation
(by definition), while in the lower panels the open circles show the
DMO simulations.  Results are shown versus halo mass (right), stellar
mass (middle), and stellar-to-halo mass ratio (left). The latter has
been shown to be better correlated with the halo response
\citep{DiCintio14, Dutton16b, Bullock17} in simulations with high star
formation thresholds, than either the stellar mass or halo mass alone.

At the lowest $\Mstar/M_{200} \sim 10^{-4}$ all of the simulations
result in no significant changes to the density profile. At higher
efficiencies of $10^{-3}$ to $10^{-2}$, the $n=0.1$ simulations still have
no significant change, but the $n=1$ simulations have a small amount
of expansion, while the $n=10$ simulations have a large amount of
expansion. At $\Mstar/M_{200} > 10^{-2}$, the $n=0.1$ and $n=1$
simulations result in halo contraction, which strengthens for higher
efficiencies. At the highest $\Mstar/M_{200}$, the $n=10$ simulations
also contract, but not as strongly as the lower $n$ simulations.

We note that our results for $n=0.1$ are very similar to those for the
APOSTLE and AURIGA simulations recently presented by \citet{Bose18}.
This is in spite of numerous differences between the codes, including
how supernova feedback is modeled and the hydrodynamical
schemes. This strengthens the notion that the star formation threshold
is the key sub-grid parameter that controls the halo response in
haloes of mass $M_{200}\lta 10^{12}\Msun$.

%% FIGURE 4
%------------------------------------------
\begin{figure*}
  \includegraphics[width=0.85\textwidth]{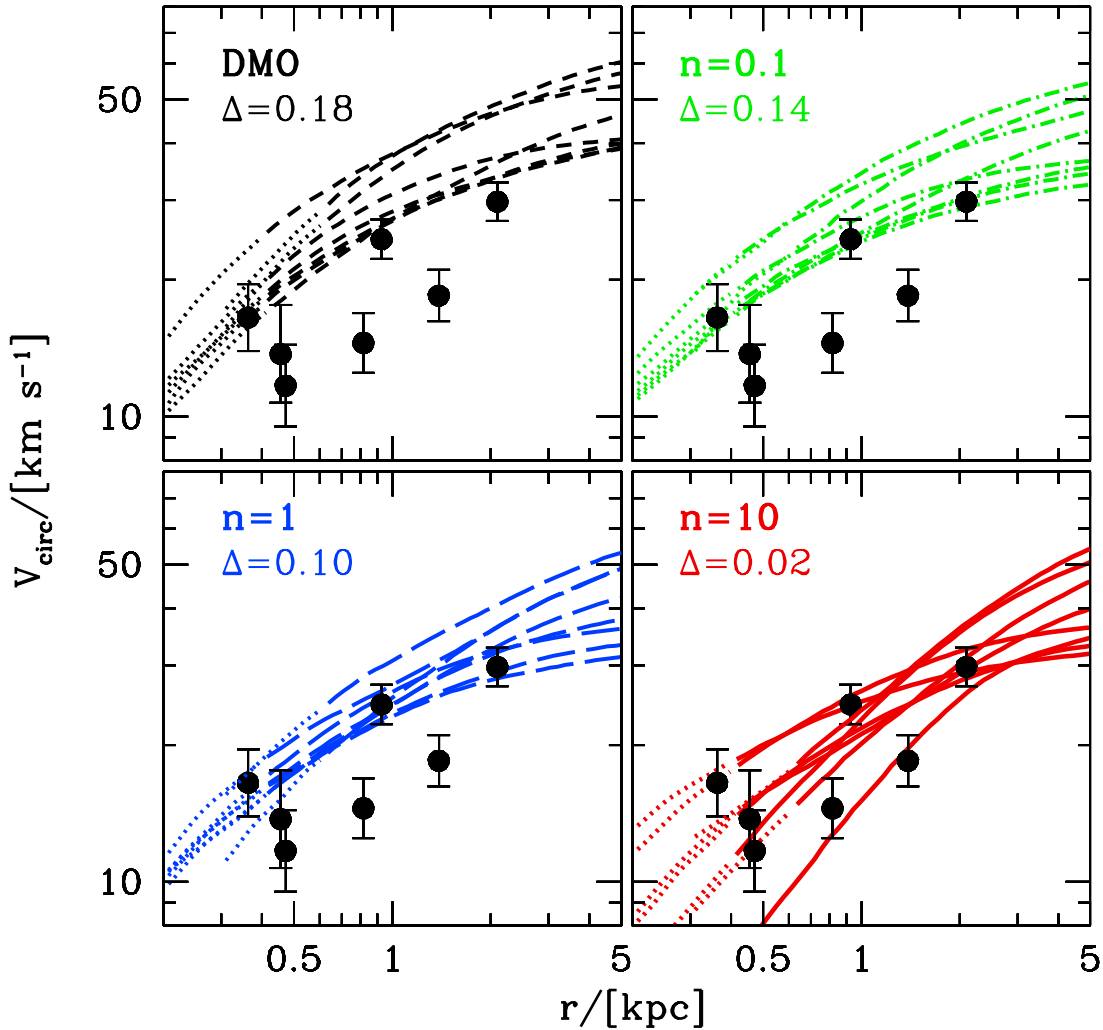}
  \caption{Circular velocity versus radius. Lines show simulations
    with stellar masses $10^6 \lta \Mstar \lta 10^8\Msun$. The
    transition from dotted to solid/dashed lines marks the scale that
    is accurately resolved (twice the dark matter softening).  Points
    with error bars show observed field galaxies more than 500 kpc
    from the Milky Way \citep{Kirby14}. The parameter $\Delta$ is the
    mean offset between the observations and the simulations. The DMO
    simulations (upper left) are offset from the observations by an
    average of 0.18 dex (i.e., a factor of 1.5). As the star formation
    threshold increases the offset decreases, such that with $n=10$
    (lower right) the offset is just 0.02 dex. This shows that
    different star formation thresholds result in testable differences
    in the structure of cold dark matter haloes.}
  \label{fig:tbtf}
\end{figure*}
%------------------------------------------

%% FIGURE 5
%------------------------------------------
\begin{figure*}
  \includegraphics[width=0.85\textwidth]{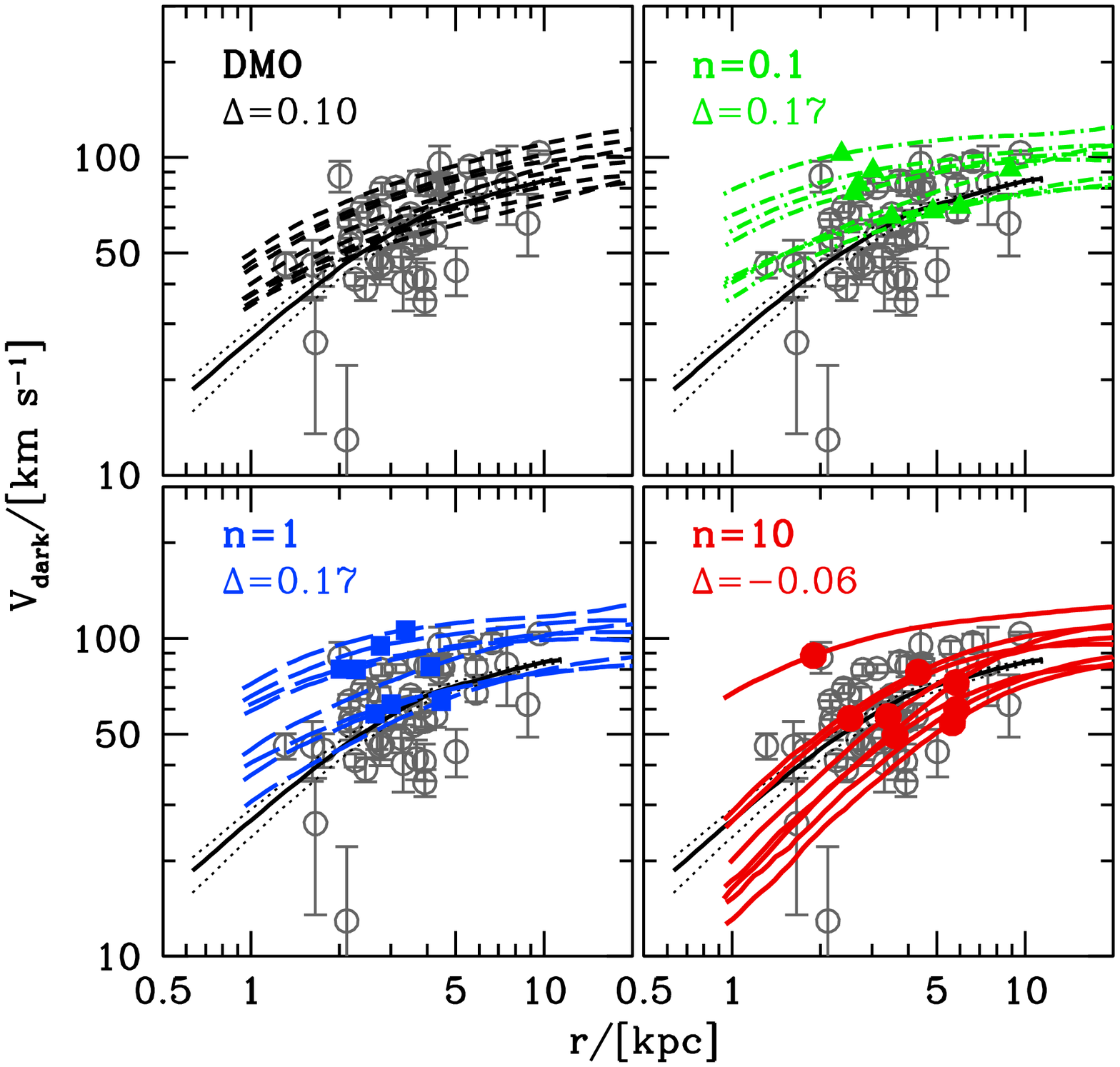}
  \caption{Dark matter circular velocity versus radius for galaxies
    with $10^9 \lta \Mstar \lta 10^{10} \Msun$.  The same observations
    are shown in all panels. Grey circles with error-bars show the
    dark matter circular velocity within the half-light radius for
    observed galaxies from SPARC.  The    solid black line shows the
    mean dark matter circular velocity curve of the observations,
    while the dotted lines show the impact of a 0.1 dex uncertainty in
    stellar mass-to-light ratio.  Each panel shows a different set of
    simulations: DMO (upper left, black); $n=0.1$ (upper right,
    green); $n=1.0$ (lower left, blue); and $n=10$ (lower right,
    red). The points correspond to the half stellar mass radii .  The
    value $\Delta$ is the mean offset [dex] between the simulations
    and the observed dark matter velocity at 2 kpc.}
  \label{fig:vr_sparc}
\end{figure*}
%------------------------------------------

%% SECTION 4
\section{Velocity profiles}

We now discuss the circular velocity profiles of the simulations to
highlight the magnitude of the observational differences one can
expect. We split the simulations into three mass ranges corresponding
to dwarf galaxies ($10^6 \lta \Mstar/\Msun \lta 10^8$), intermediate
mass galaxies ($10^9 \lta \Mstar/\Msun \lta 10^{10}$), and Milky Way
mass galaxies ($\Mstar \sim 5\times 10^{10}\Msun$).

\subsection{Dwarf galaxies}

We start with the 8 lowest mass haloes which form galaxies of stellar
masses $10^6 \lta \Mstar \lta 10^8 \Msun$.  Fig.~\ref{fig:tbtf} shows
the circular velocity profiles (lines) of these simulations compared
to the circular velocity at the 3D half-light radius of field dwarf
galaxies in the local group (points with error bars) from
\citet{Kirby14}.  We have selected observed dwarfs with V-band
luminosities from $10^6$ to $2\times 10^8 \Lsun$ with distances of at
least 500 kpc ($\sim 2$ virial radii) from the Milky Way to minimize
contamination of back-splash galaxies \citep{Buck18}.

Qualitatively, we see that the DMO simulations are systematically too
high, while the $n=10$ simulations can match all of the observations.
The lower threshold simulations ($n=0.1, n=1$) can reproduce some, but
not all of the observed data points.  To be more quantitative, we
calculate the average offset between the observations ($V_{\rm obs}$)
and simulations $V_{\rm sim}$.  For each observed data point, $V_{{\rm
    obs},i}$, the mean offset with respect to the $N_{\rm sim}=8$
simulations is
\begin{equation}
\Delta_i=\sum_{j=1}^{N_{\rm sim}}(\log_{10} V_{{\rm obs},i} -\log_{10} V_{{\rm sim},j})/N_{\rm sim}.
\end{equation}
We then take the mean of $\Delta_i$ over the 7 observed data points,
which we denote $\Delta$. For DMO (upper left panel), $\Delta=0.18$,
i.e. the average offset between simulation and observation is a factor
of 1.5 in velocity, and a factor of 2.3 in enclosed mass.  This
recovers the well known Too-big-to-fail (TBTF) problem of local group
field galaxies \citep{Garrison-Kimmel14}.  We showed previously in
\citet{Dutton16a} that the NIHAO simulations resolve this
problem. Indeed for the fiducial NIHAO $n=10$ (lower right), the
simulations match the observations well with $\Delta=0.02$. However,
the hydro simulations with lower thresholds predict less halo
expansion and tend to over-predict the observations. The $n=1$
simulations have $\Delta=0.10$ (a factor of $1.25$ to high), and the
$n=0.1$ simulations have $\Delta=0.14$ (a factor of $1.4$ to high).

%% FIGURE 6
%------------------------------------------
\begin{figure*}
  \includegraphics[width=0.85\textwidth]{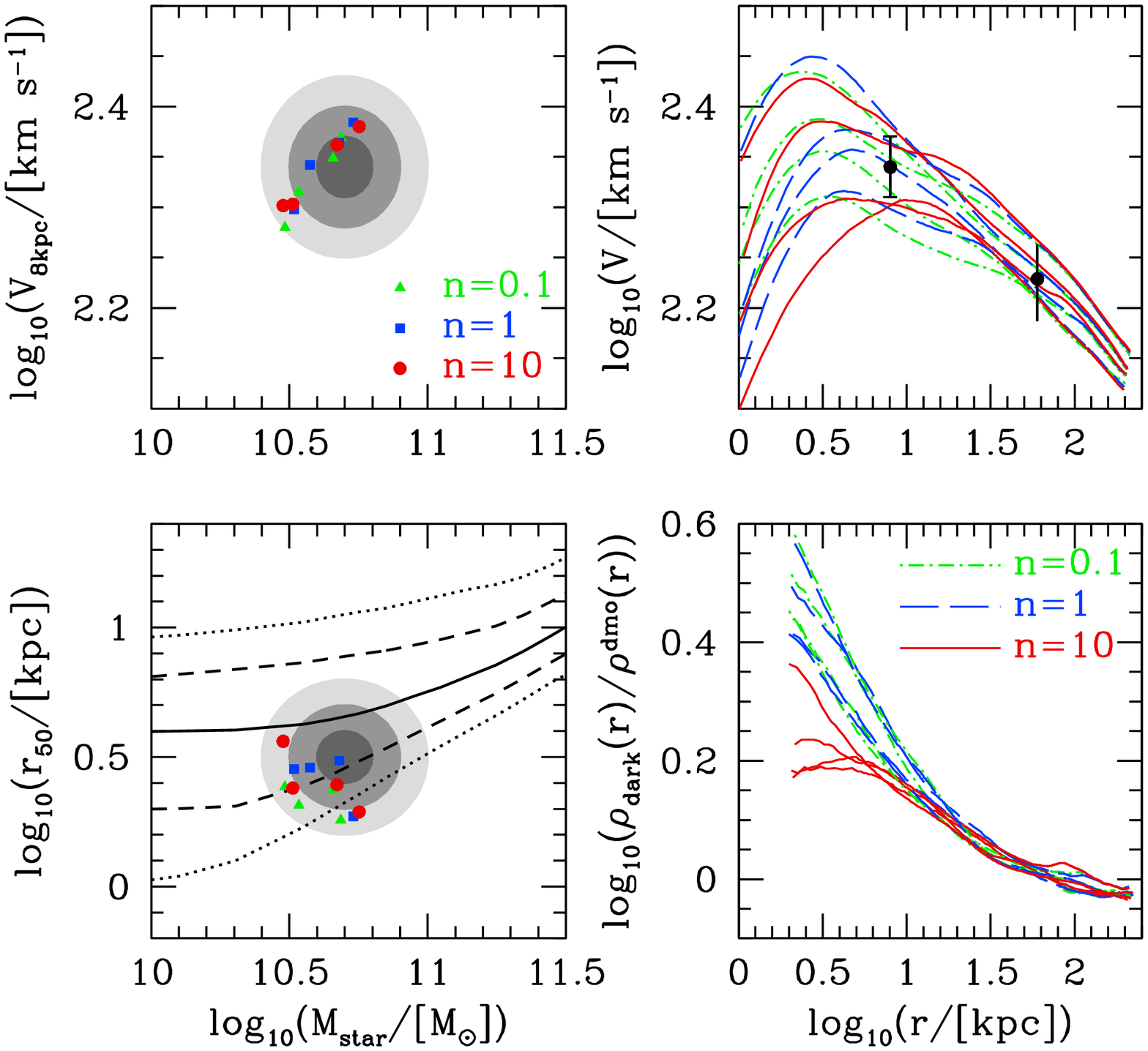}
  \caption{Results for Milky Way mass haloes. Left panels show the
    circular velocity at 8 kpc versus stellar mass (upper) and projected
    half stellar mass radius versus stellar mass (lower). Ellipses show
    the 1,2, and $3\sigma$ uncertainties for the Milky Way. For the
    sizes the lines show the median (solid), $1\sigma$ (dashed) and
    $2\sigma$ (dotted) relations for observations from SDSS. The upper
    right panel shows the total circular velocity versus radius. The
    points with error bars show observational constraints. The lower
    right panel shows the change in enclosed dark matter density
    profile between hydro and DMO simulations. In all panels the colour
    refers to the star formation threshold: $n=10$ (red), $n=1$
    (blue), $n=0.1$ (green).}
  \label{fig:mw}
\end{figure*}
%------------------------------------------

The current observations clearly favor CDM simulations with a high
star formation threshold. Turning this around, if a low star formation
threshold can be shown to be a better description for star formation
in sub-grid models of galaxy formation, then this test can be used to
falsify the CDM model.  While this test is not conclusive due to the
small number statistics of the observations, we can conclude that
simulations with different star formation thresholds make clear
testable differences in the circular velocities on scales of 1 kpc.

\subsection{Intermediate mass galaxies}

We next consider galaxies of stellar mass $10^9 \lta \Mstar \lta
10^{10}\Msun$. There are 8 sets of simulations in this mass range.
These simulations show a large disparity in the halo response with
strong expansion for $n=10$ and no change or mild contraction for
$n=0.1$ and $n=1$ (see Fig.~\ref{fig:alpha}).

We compare to galaxies from the SPARC survey of nearby star forming
galaxies \citep{Lelli16}.  Fig.~\ref{fig:vr_sparc} shows the dark
matter circular velocity profiles. For observations these are
obtained by subtracting the stellar and gas circular velocity
profiles from the total rotation velocity, assuming a stellar
mass-to-light ratio at $3.6\,\mu$m of 0.5. For the observations the
grey symbols show the dark matter velocity at the half-light radius.
The solid black line shows the average dark matter velocity profile of
the observations plotted between the average smallest and largest
point on the rotation curve. Because these galaxies tend to be dark
matter dominated, there is only a small uncertainty in the dark matter
profile caused by the 0.1 dex uncertainty in stellar mass-to-light
ratio (dotted lines).

For the NIHAO simulations, each panel shows a different simulation:
DMO (top left), $n=0.1$ (top right), $n=1$ (bottom left), and $n=10$
(bottom right). The lines show the dark matter circular velocity
profiles, where the DMO has been rescaled by the cosmic baryon
fraction ($\sqrt{1-\fbar}\simeq 0.92$).  Symbols are located at the
projected half-mass radius of the stars. This shows that the galaxy
sizes for these simulations are in reasonable agreement with the
observations and that there is only a small dependence of the sizes on
the star formation threshold.  Overall, every simulated galaxy has an
observed counterpart within a small interval of radius and velocity.
Going further, we can match the full circular velocity curves, or dark
matter circular velocity curves, to find good observational analogs of
all the simulated galaxies. 

However, only the $n=10$ simulations can reproduce the full range of
observed velocities in Fig.~\ref{fig:vr_sparc}. The low thresholds are
unable to significantly expand the dark matter halo.  Qualitatively
when comparing the simulated dark matter profiles to the observed mean,
we see that the high threshold ($n=10$) simulations tend to be below
the observed mean, while the low threshold simulations ($n=0.1, n=1$)
tend to be above the mean relation.  Being quantitative, the parameter
$\Delta$ is the mean offset between the simulations and the observed
dark matter velocity at 2 kpc. DMO has $\Delta=0.1$. The $n=0.1$ and
$n=1$ simulations have higher $\Delta=0.17$, which indicates the haloes are
contracting. The $n=10$ simulations have $\Delta=-0.06$
indicating expansion.  Our sample of simulations is small, so the
cosmic variance could be large.  With that caveat aside, there are two
generic solutions to resolve this discrepancy.

On the observational side there could be a systematic that biases the
rotation curves low (or high), possibly from non-circular motions due
to pressure support or triaxial dark matter haloes
\citep{Valenzuela07, Oman19}.  On the theoretical side, a threshold between 1
and 10 could possibly result in a better match to observations.  More
interesting is the possibility that a single star formation threshold
is too simplistic and that a variable threshold captures better the
physics of star formation at the resolution of our simulations. This
variability could occur systematically with other properties of the
gas, or it could be essentially a random variable at the scales we are
considering. Indeed \citet{Semenov17} present a model in which the
threshold for star formation is a function of both the gas density,
$n$, and the total subgrid velocity dispersion, $\sigma_{\rm tot} =
\sqrt{c_s^2 + \sigma_{\rm turb}^2}$, where $c_s$ is the sound speed,
and $\sigma_{\rm turb}$ are the explicitly modeled sub-grid turbulent
velocities. In this model the effective star formation threshold
varies from $n \sim 10$ to $n \sim 1000$.

\subsection{Milky Way mass galaxies}

Fig.~\ref{fig:mw} shows results for the four most massive simulations.
Top left shows circular velocity at 8 kpc versus stellar mass, bottom
left shows projected half stellar mass radius versus stellar
mass. Observational estimates for these values for the Milky Way are
shown with ellipses corresponding to the 1,2,3 $\sigma$ uncertainty
which are based on results from dynamical models of \citet{Widrow08}
and \citet{Bovy13}.

\begin{itemize}
\item {\bf Circular Velocity.} We adopt a circular velocity at 8 kpc
  (the solar radius) of $\log_{10}(V_{8 \rm kpc}/[\kms]) = 2.34\pm
  0.03$, which gives a $2\sigma$ range from 191 to 251 $\kms$.

\item {\bf Stellar Mass.} We adopt $\log_{10}(M_{\rm
  star}/[\Msun])=10.7 \pm 0.1$, which gives a $2\sigma$ range from 3.2
  to $7.9\times 10^{10}\Msun$.  \citet{Widrow08} finds a disk stellar
  mass of $4.22\pm0.51 \times 10^{10}\Msun$ and a bulge mass of
  $0.96\pm0.12 \times 10^{10}\Msun$. \citet{Bovy13} finds a disk
  stellar mass of $4.6\pm0.3\times 10^{10}\Msun$. 

\item {\bf Half-Mass Size.} We adopt $\log_{10} (r_{50}/[{\rm
    kpc}])=0.5\pm0.1$, which gives a $2\sigma$ range  from 2.0 to 5.0
  ${\rm kpc}$.  \citet{Widrow08} finds a disk scale length of $R_{\rm
  d}=2.8\pm0.23$ (their prior had limits 2.0 to 3.8 kpc),
  corresponding to a half-mass size of $r_{50}=4.7\pm0.4$.
  \citet{Bovy13} finds a disk stellar mass scale length of $R_{\rm
    d}=2.15\pm0.14$ kpc, corresponding to a half-mass size of
  $r_{50}=3.61\pm0.23$. These disk sizes are likely an upper limit to
  the total half-light size once the central bulge is included. For
  example, if 20 per cent of the stellar mass is in a compact bulge
  \citep{Widrow08}, then the half-mass radius of the galaxy is roughly
  1.1 disk scale lengths.  For the sizes the lines show observations
  from SDSS \citep{Dutton11,Simard11} for various percentiles of the
  distribution of sizes in bins of stellar mass: median (solid), 15.9
  and 84.1 (dashed), 2.3 and 97.7 (dotted).  This shows that the Milky
  Way is about $1\sigma$ smaller (with a large uncertainty) than
  typical galaxies of the same stellar mass.
\end{itemize}

We see that the velocities, stellar masses, and sizes are not
sensitive to the star formation threshold and that all 12 simulations
fall within the observed $3\sigma$ ranges.

The top right panel shows the total circular velocity profile from 1
kpc to the virial radius.  The circular velocity profile has two
observational data points (black error bars): the velocity at 8 kpc
(from above), and the velocity at 60 kpc \citep{Xue08}. All simulations
are consistent with these observational constraints. 
The bottom right panel shows the change in the dark matter
density profile with respect to DMO.  We see similarities and
differences in halo response between simulations with different star
formation thresholds.

Similarities: At large radii ($\gta 20$ kpc) the profiles are
indistinguishable. At smaller radii all haloes contract, with more
contraction at smaller radii.  These results are qualitatively similar
to previous studies of Milky Way mass haloes with both low ($n\sim
0.1$) \citep{Marinacci14} and high ($n\sim 100$) \citep{Chan15} star
formation thresholds. 

Differences: At radii below $\sim 10$ kpc the high threshold
simulations (red lines) result in less contraction than the low
threshold simulations. In three of the four $n=10$ simulations, the
change in density profile levels off (indicating a similar asymptotic
inner slope as the DMO), whereas in all of the $n=1$ and $n=0.1$
simulations the contraction is larger at smaller radii (indicating a
steeper asymptotic slope than DMO).  These differences have
implications for the dark matter annihilation signal from the galactic
center, since the signal goes as density squared.

%% SECTION 5
\section{Physical mechanism for halo expansion}

We have shown how the structure of the dark matter halo depends
strongly on the star formation threshold used in the simulation.  We
now discuss the physical mechanism that is driving different halo
responses and observational ways of distinguishing between them.

Looking at movies of the evolution of the gas and stars, we see that
there are clear differences in the spatial and temporal distribution
of the star formation. Specifically, higher thresholds result in more
bursty star formation (i.e., concentrated in space and time), while
lower thresholds result in more uniform distributions of star
formation in both space and time. 

Gas flows driven by SN feedback cause halo expansion when they result
in rapid variability of the potential \citep{Pontzen12}.  In this
case, rapid means fast compared to the dynamical time. Changes in
particle orbits go like the change in the potential squared
\citep{Pontzen12}.  Using a toy model for adiabatic inflows followed
by impulsive outflows \citet{Dutton16b} showed that the ratio between
the final and initial radius of a shell of dark matter goes as $r_{\rm
  f}/r_{\rm i} = 1+f^2$, where $f$ is the ratio between the gas mass
removed from radius $r_{\rm i}$ to the total enclosed mass at radius
$r_{\rm i}$. Thus one outflow event of say $f=0.2$ has an order of
magnitude greater impact on the structure of the halo than 10 outflow
events of $f=0.02$, even though the total integrated outflow mass is
the same.

%% FIGURE 7
%------------------------------------------
\begin{figure}
  \includegraphics[width=0.45\textwidth]{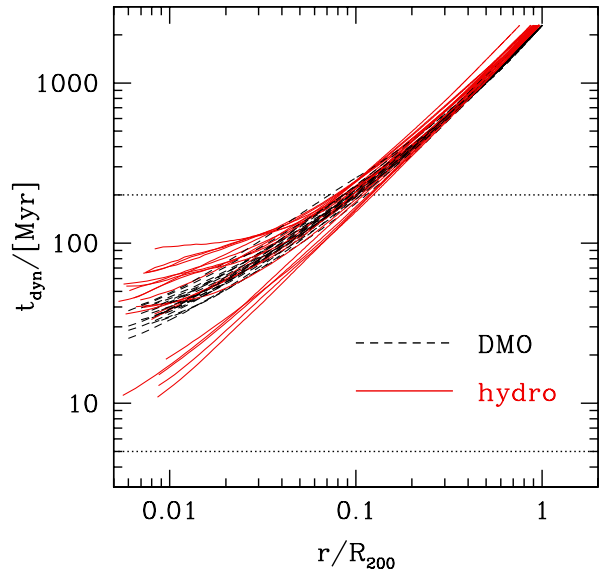}
  \caption{Dynamical time versus radius for DMO (black dashed) and hydro
    ($n=10$) simulations (red solid).  At 1 per cent of the virial radius the
    dynamical time of the DMO simulations is between 100 and 200 Myr. In
    order to expand the halo the potential fluctuations (and hence
    star formation variations) need to occur on a shorter time scale. }
  \label{fig:torb}
\end{figure}
%------------------------------------------

%% FIGURE 8
%------------------------------------------
\begin{figure*}
  \includegraphics[width=0.32\textwidth]{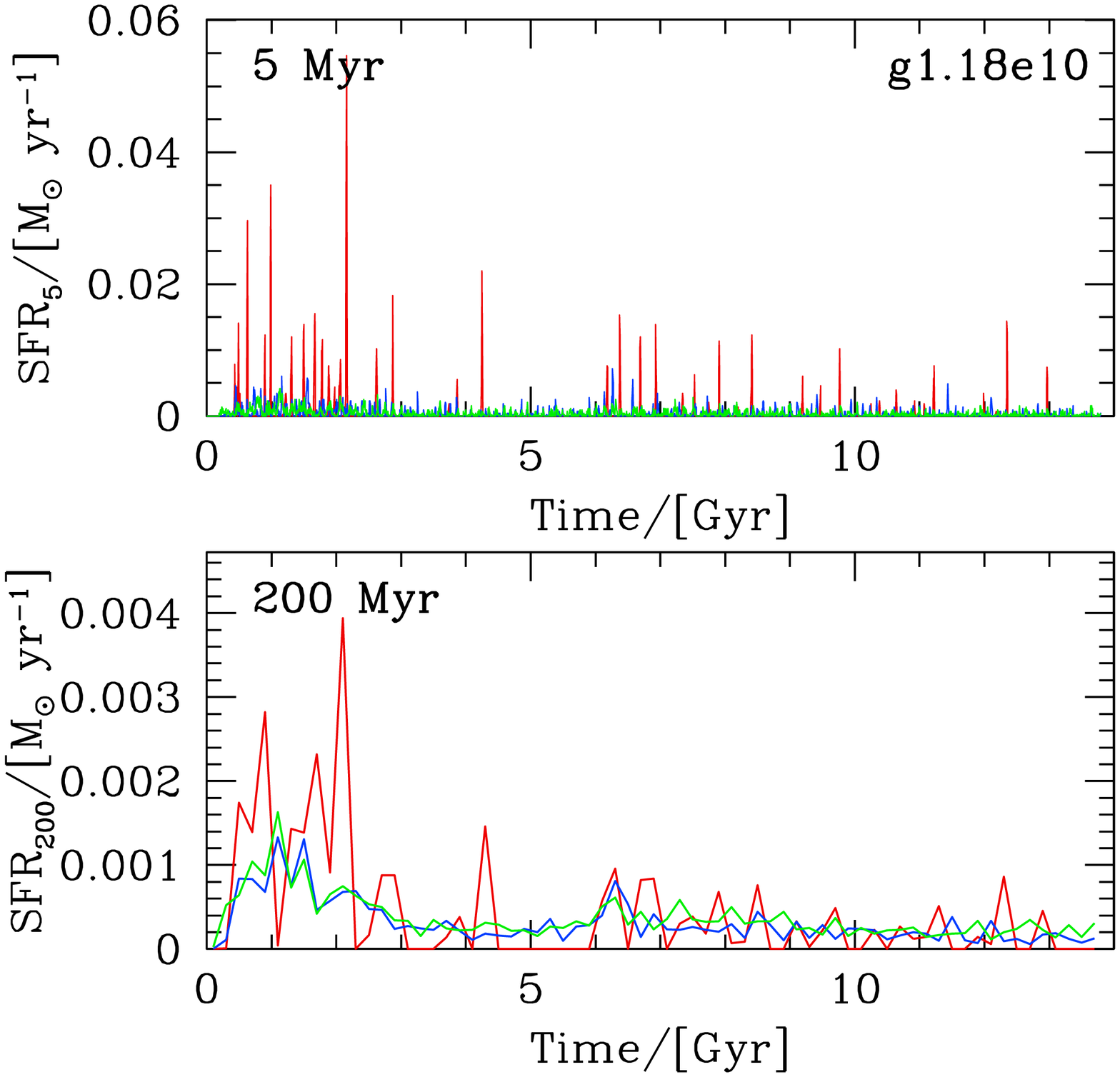}
  \includegraphics[width=0.32\textwidth]{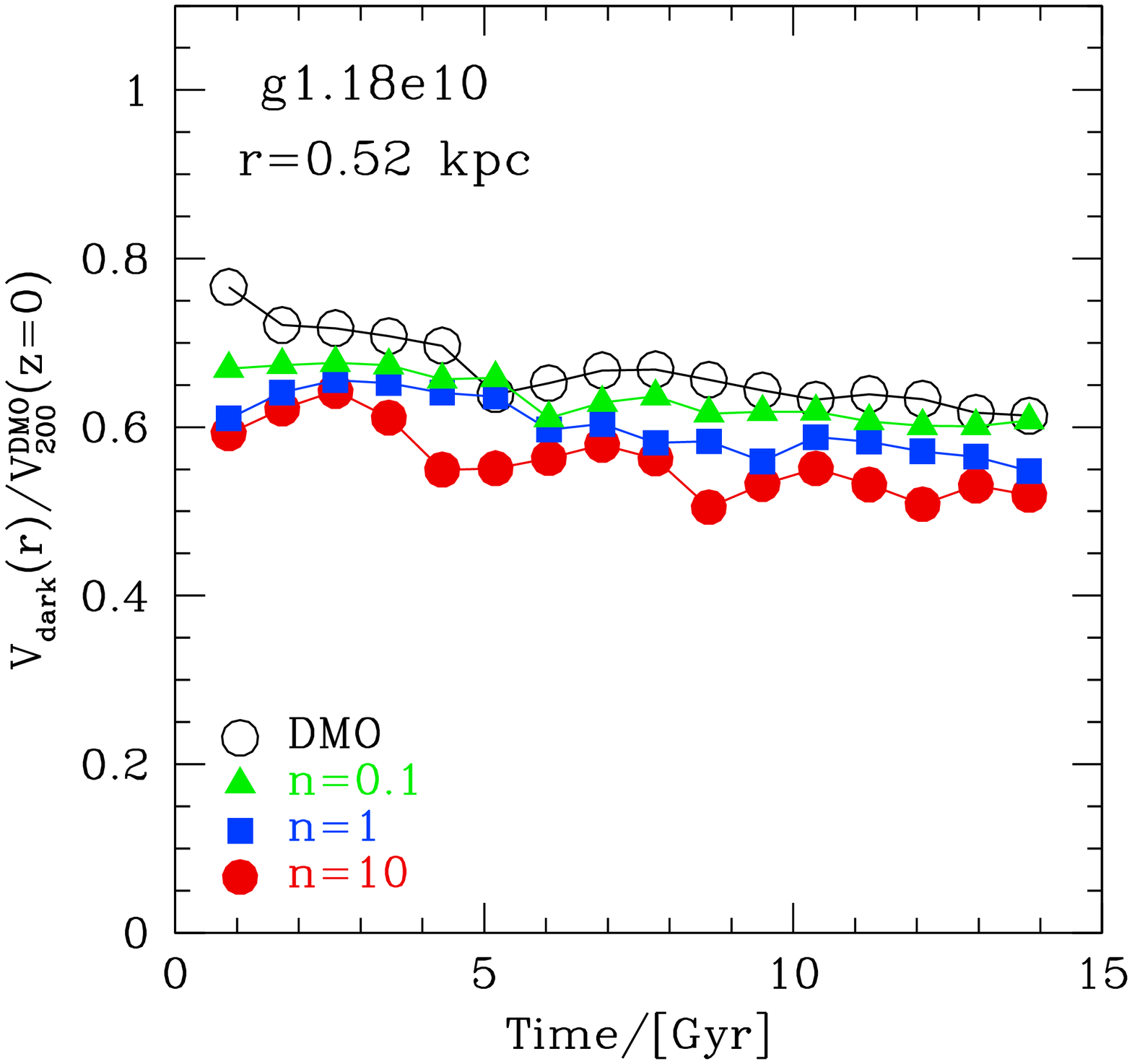}
  \includegraphics[width=0.32\textwidth]{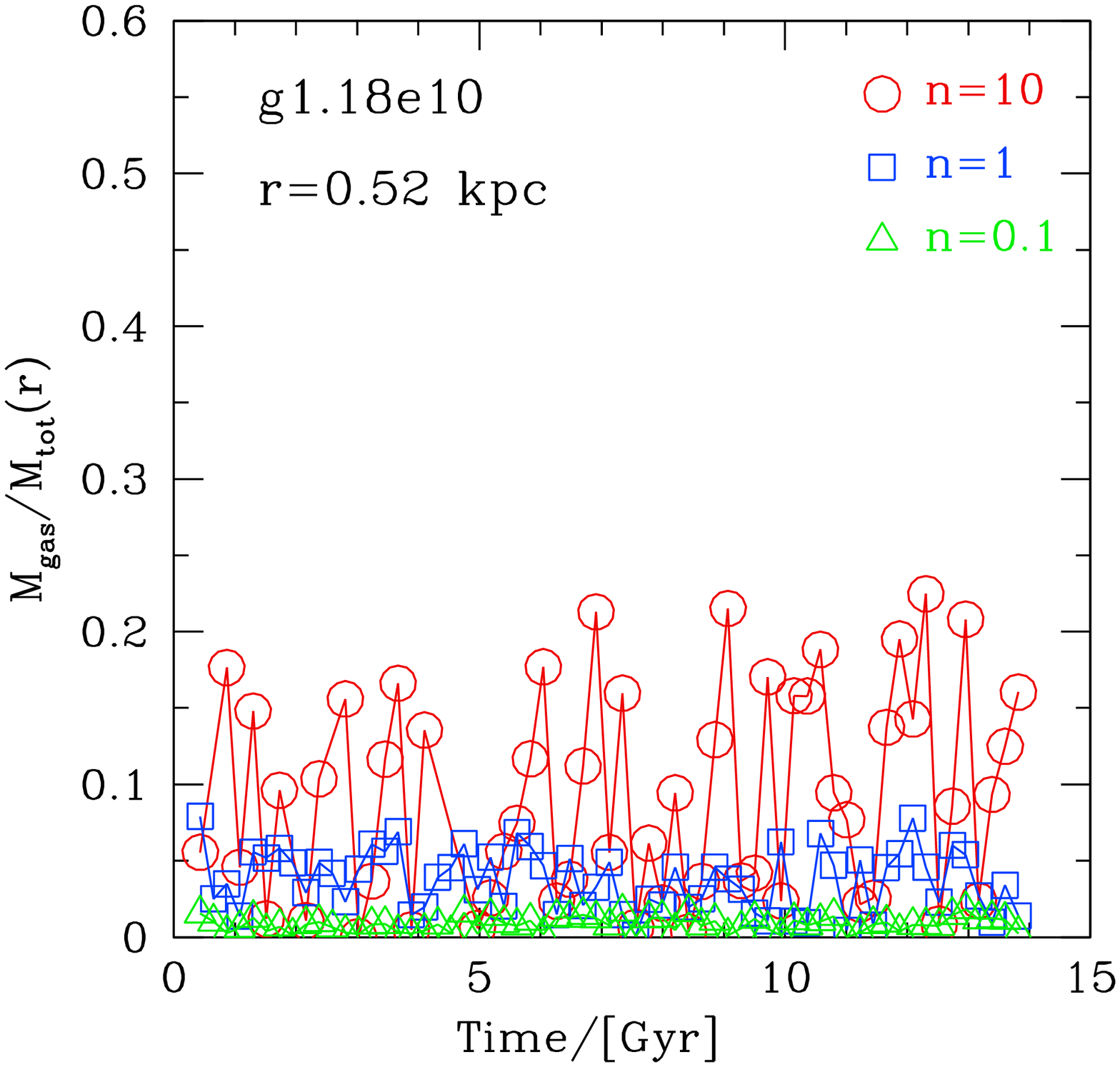}  
  \includegraphics[width=0.32\textwidth]{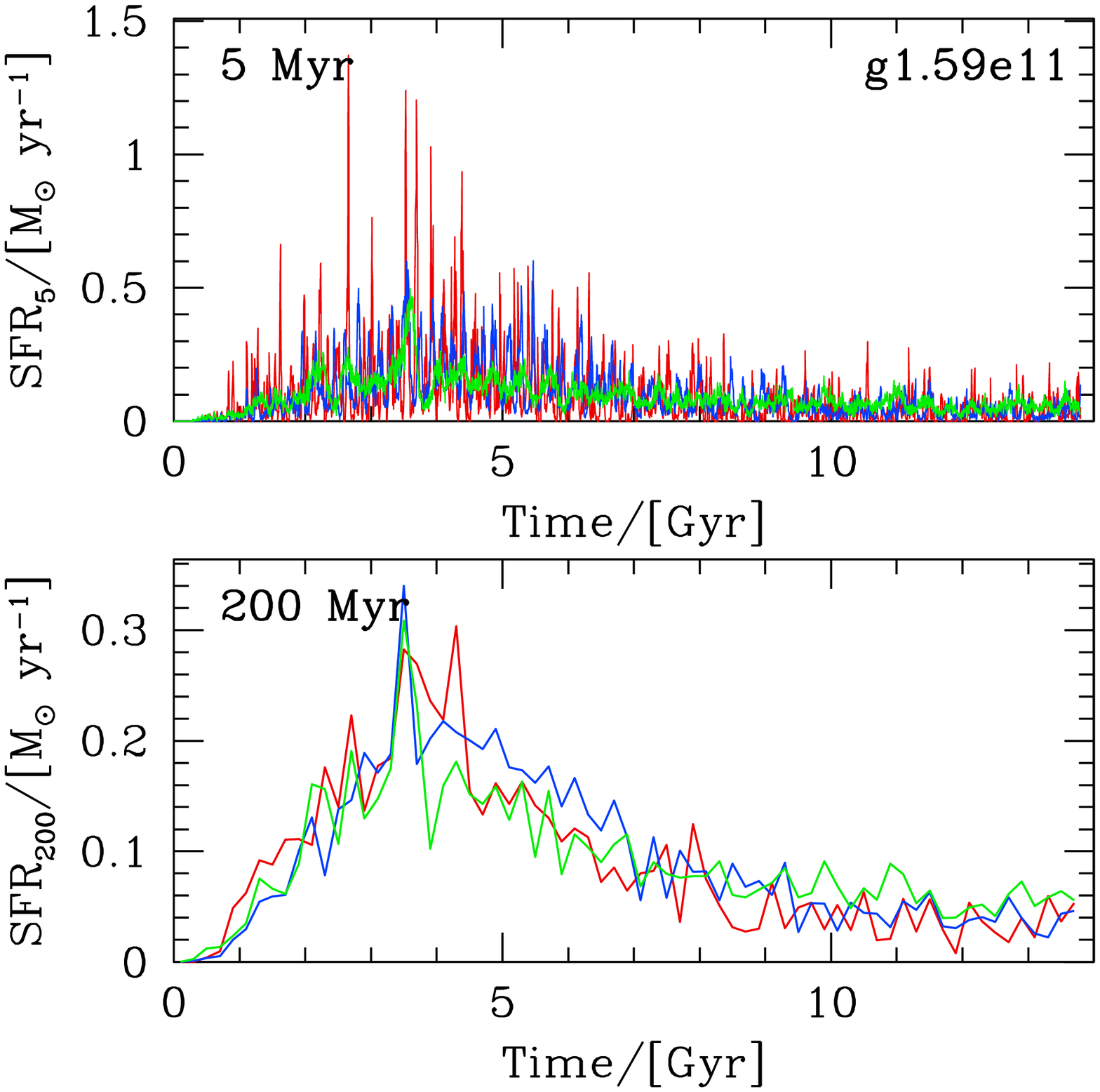}
  \includegraphics[width=0.32\textwidth]{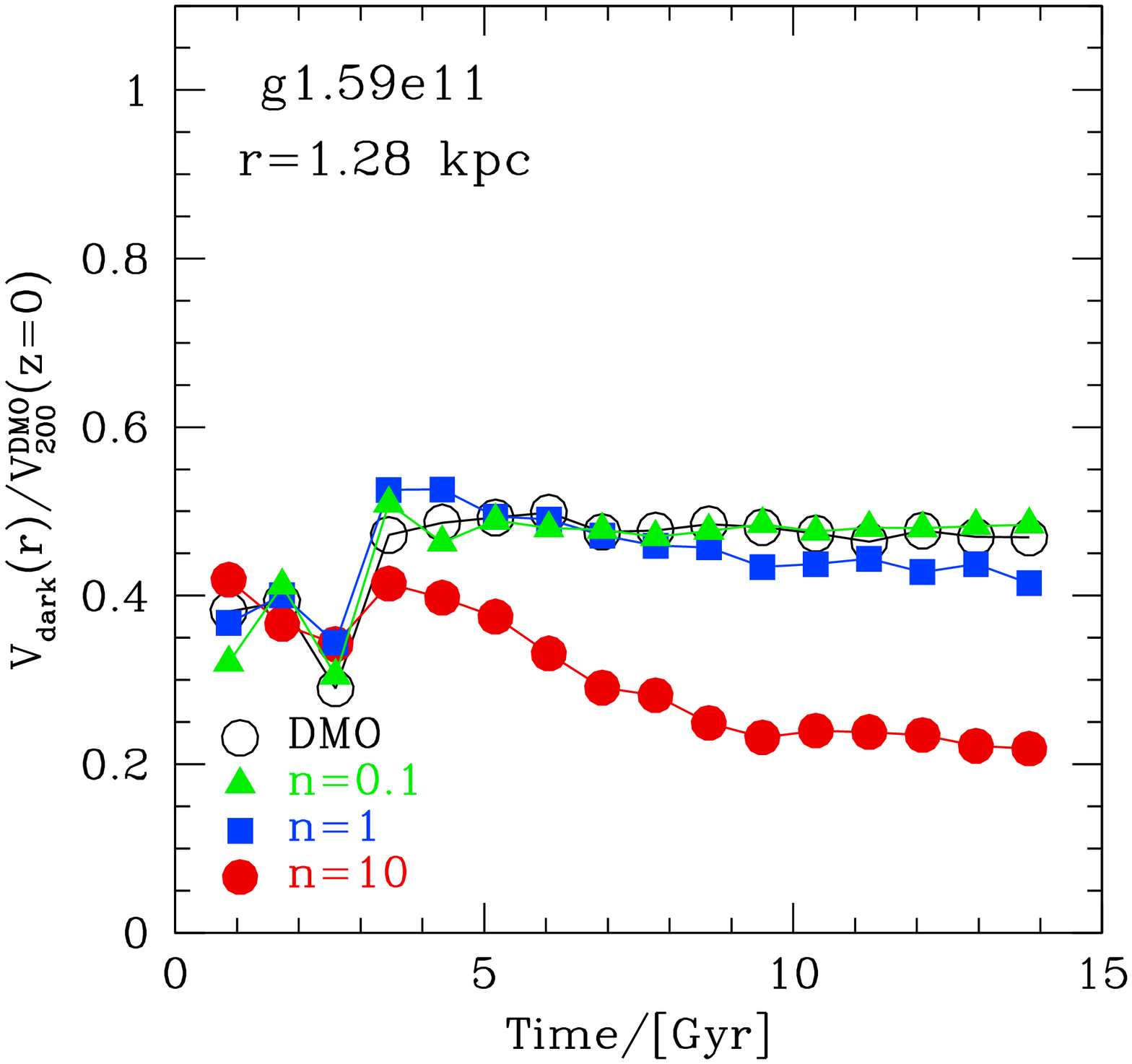}
  \includegraphics[width=0.32\textwidth]{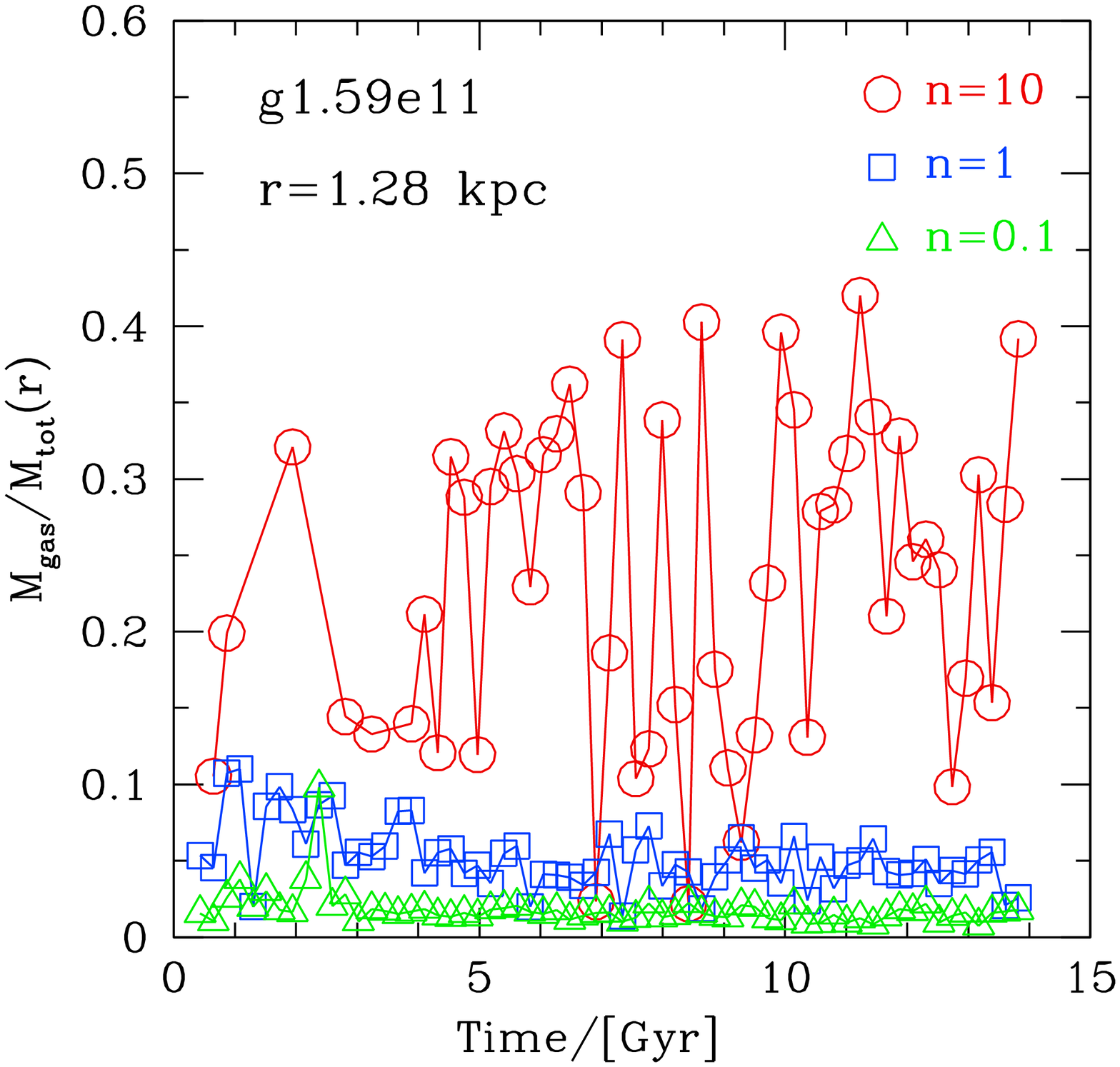}  
  \includegraphics[width=0.32\textwidth]{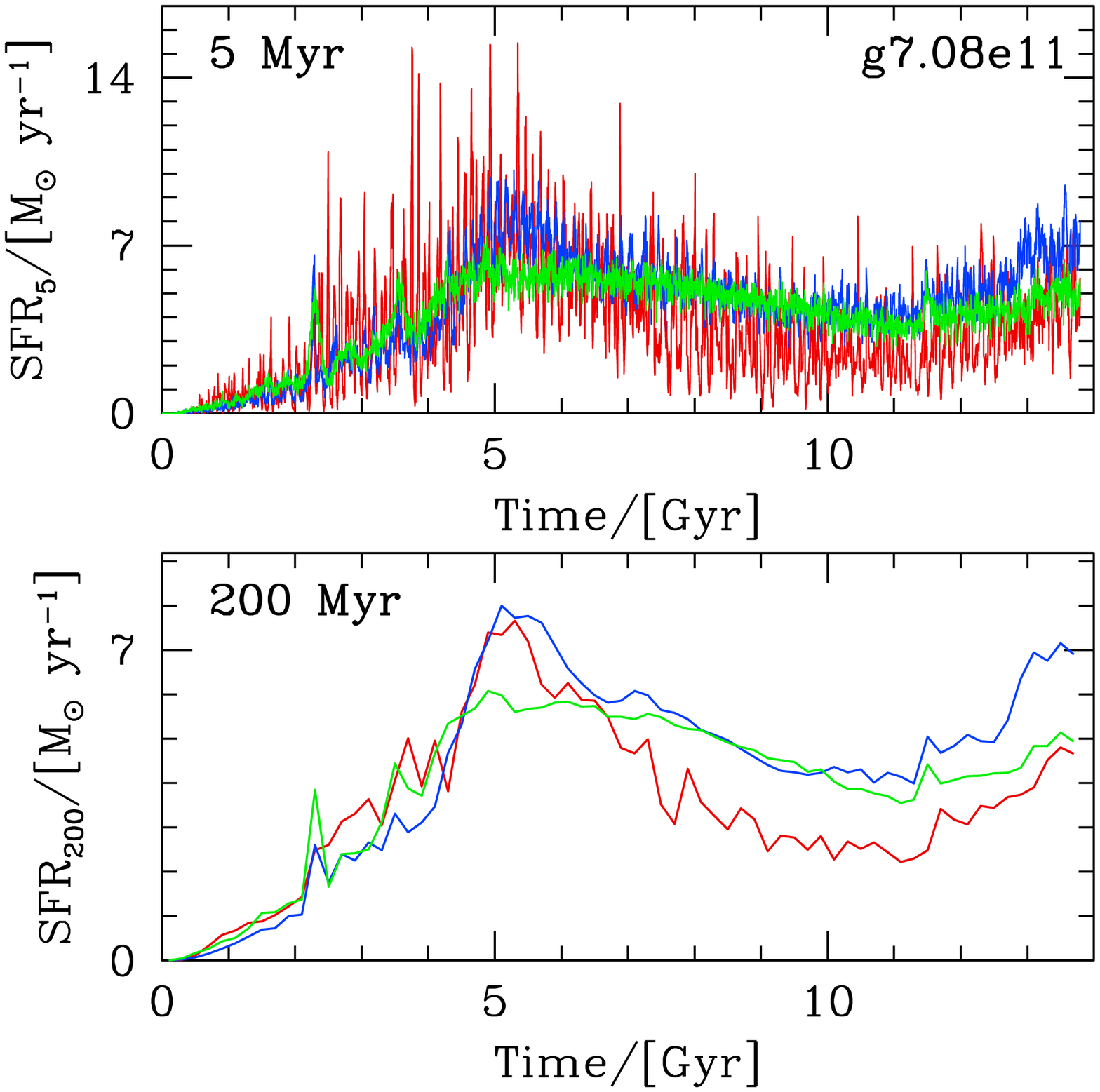}
  \includegraphics[width=0.32\textwidth]{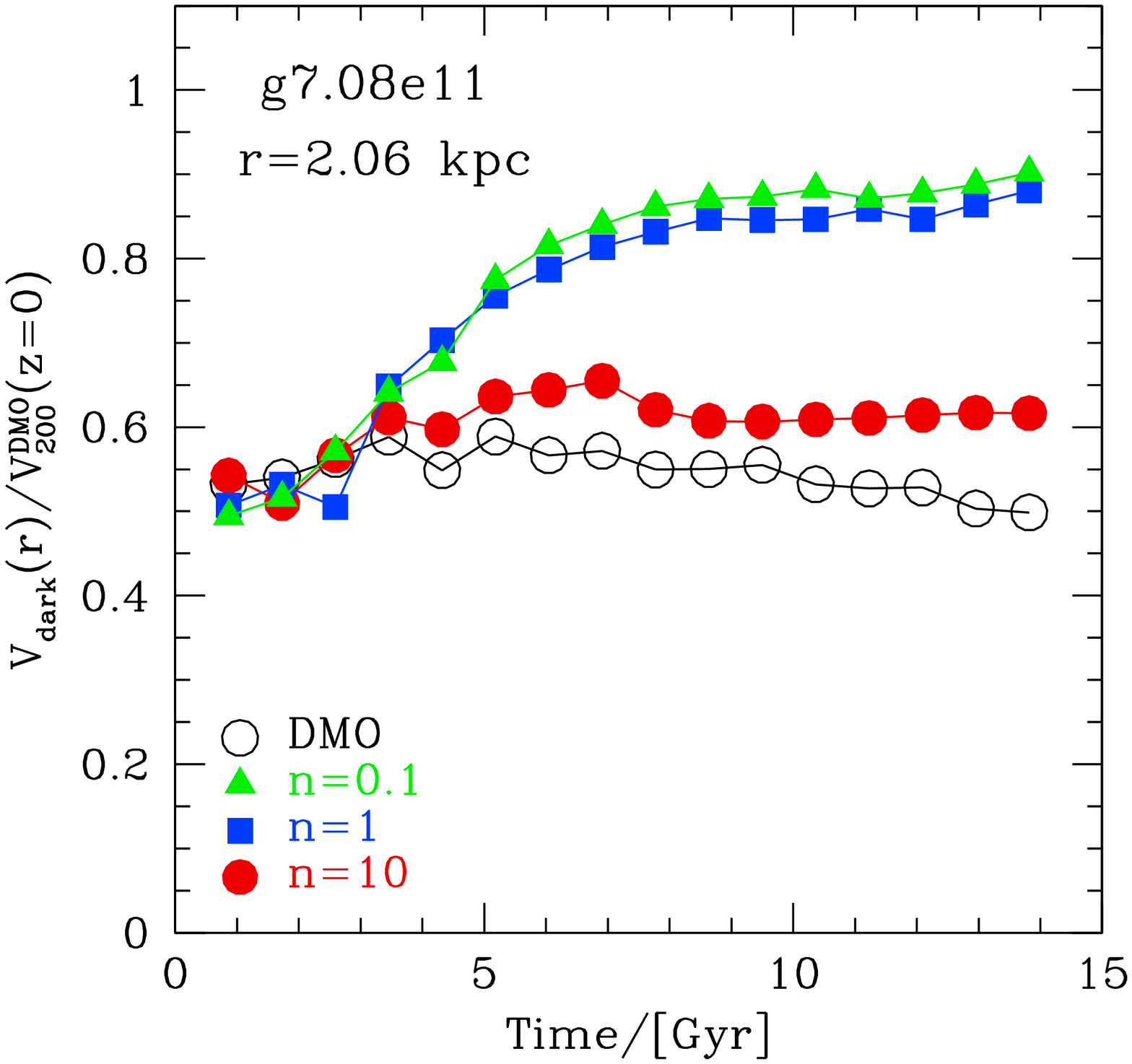}
  \includegraphics[width=0.32\textwidth]{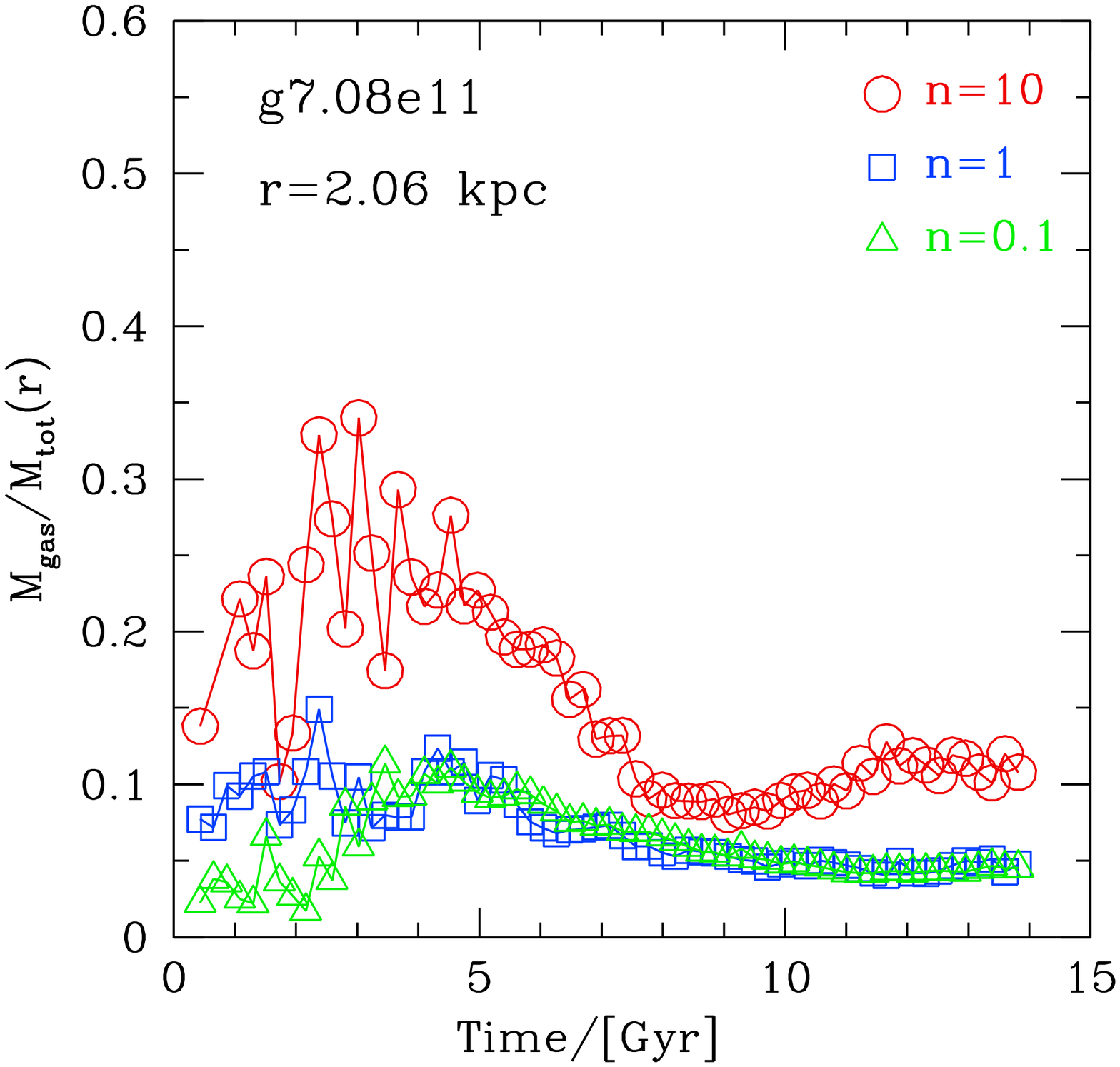}
  \caption{Star formation histories (left panels), dark matter
    circular velocities (middle panels) and gas fractions (left
    panels) for three galaxies that show the different types of halo
    response. A dwarf galaxy g1.18e10 (upper panels), an intermediate
    mass galaxy g1.59e11 (middle panels), and a milky way mass galaxy
    g7.08e11 (lower panels).  The middle panels show the change in
    dark matter circular velocity at 1 per cent of the $z=0$ virial
    radius (this radius is given under the galaxy name). For the star
    formation histories  the upper panels measure star formation in 5
    Myr intervals, while the lower panels measure in 200 Myr
    intervals. }
  \label{fig:sfh}
\end{figure*}
%------------------------------------------

%% FIGURE 9
%------------------------------------------
\begin{figure*}
  \includegraphics[width=0.45\textwidth]{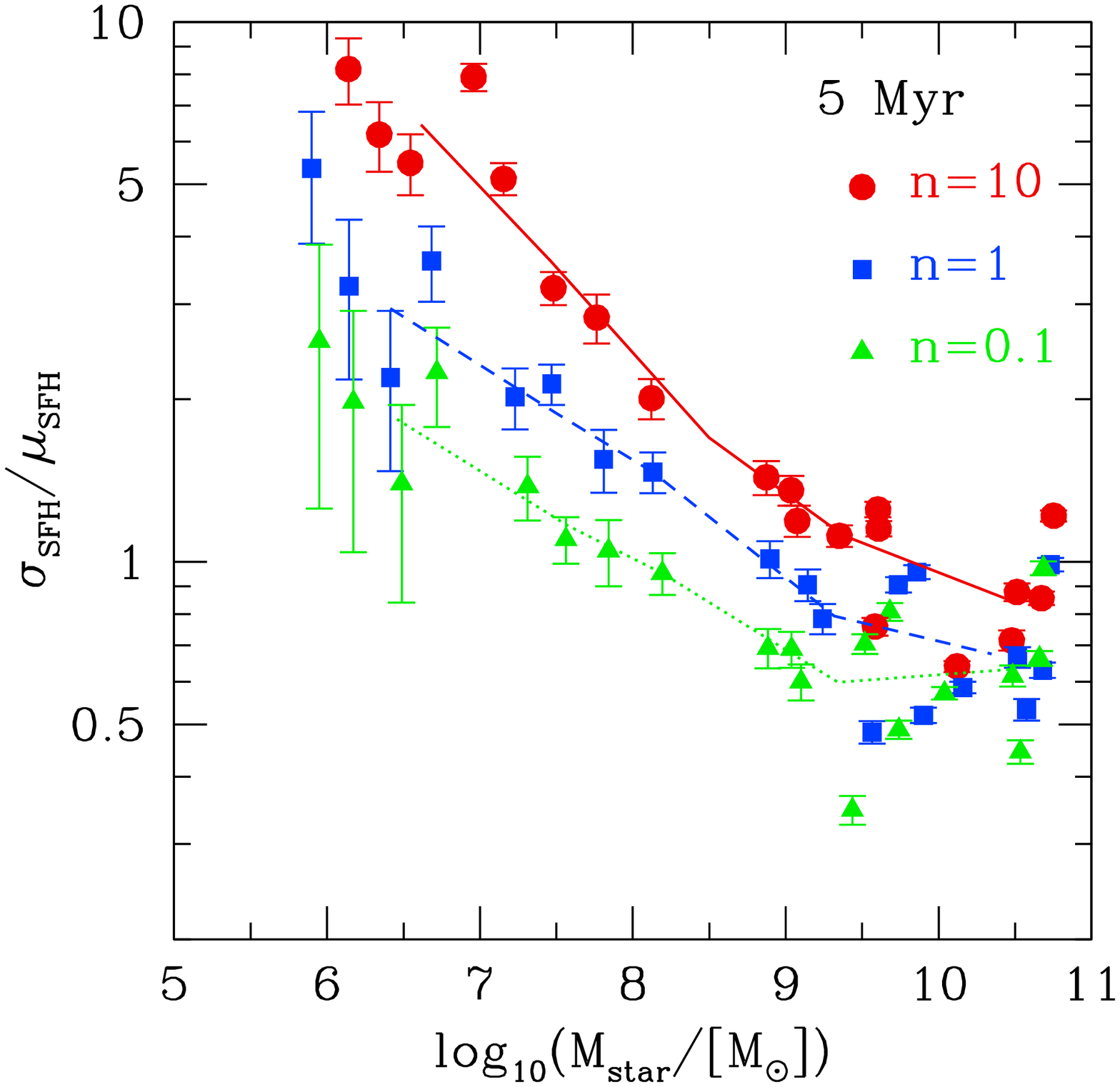}
  \includegraphics[width=0.45\textwidth]{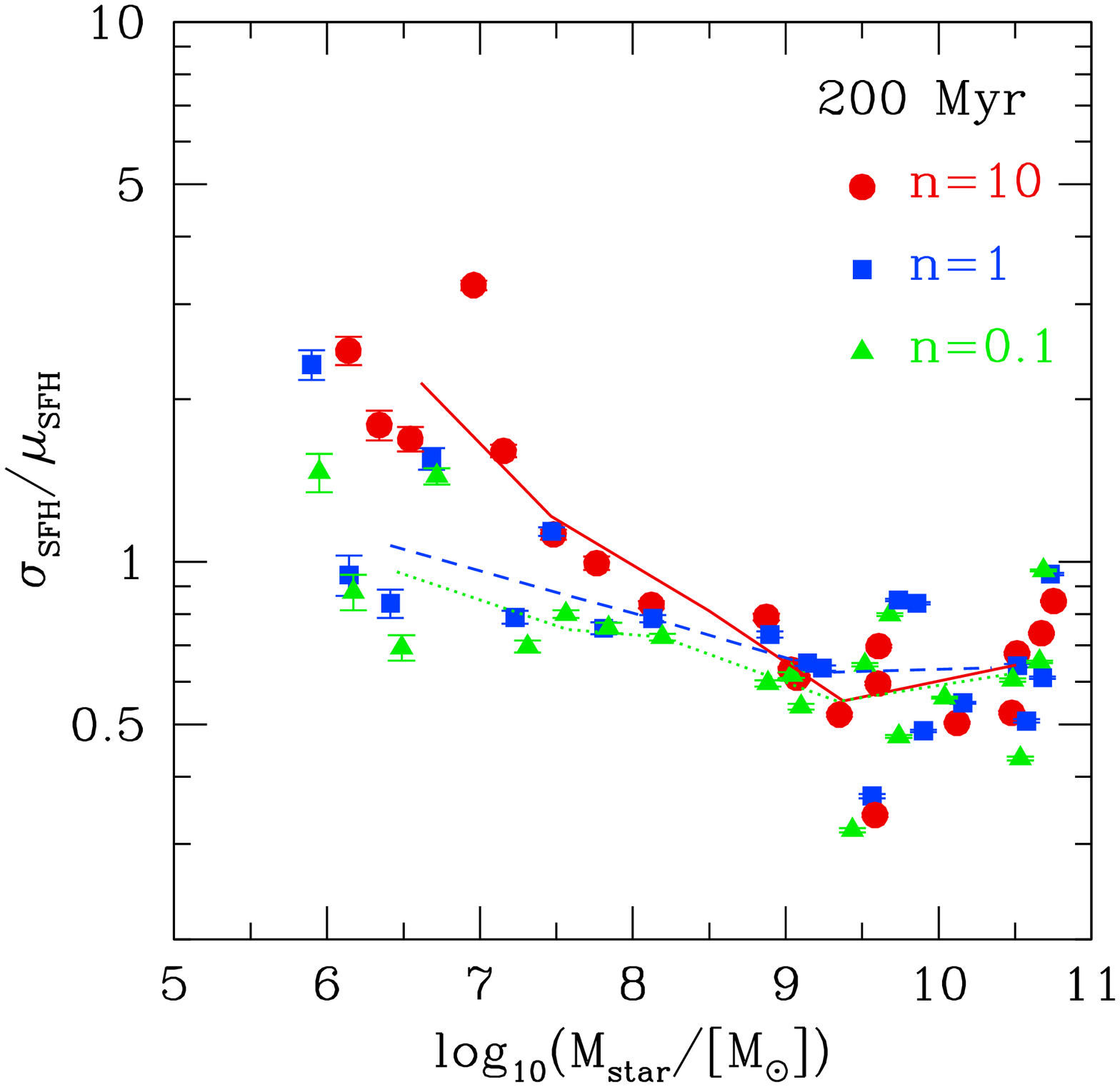}
  \caption{Variability in the star formation history versus $z=0$
    stellar mass.  The y-axis shows the ratio between standard
    deviation in SFH to the mean SFH. The error bar corresponds to the
    Poisson error on the SFR due to the number of star particles
    formed. Simulations with different star formation thresholds are
    shown as $n=10$ (red circles), $n=1$ (blue squares), and $n=0.1$
    (green triangles).  The left panel shows star formation rates
    measured over 5 Myr intervals, the right panel shows star
    formation rates measured over 200 Myr intervals. There is more
    variability in the SFH for lower mass galaxies, higher star
    formation thresholds, and shorter time scales.}
  \label{fig:sig_ssfr2}
\end{figure*}
%------------------------------------------

To get a idea of the relevant time scales, Fig.~\ref{fig:torb} shows
the dynamical time  $t_{\rm dyn} = \sqrt{ 3\pi / (16 G
  \overline{\rho})}$ versus radius for DMO simulations (black dashed
lines) and hydro simulations with $n=10$ (red solid lines). The
dynamical time  measures the time it takes to go from a radius $r$ to
$r=0$.  At 1 per cent of the viral radius the dynamical time varies
between 30 and 50 Myr for DMO simulations.  For hydro simulations,
there is more variation, because some haloes expand (increasing
$t_{\rm dyn}$) while others contract (decreasing $t_{\rm dyn}$.).
This motivates us to measure the star formation rates on a time scale
significantly less than $50 $ Myr. Variation of the star formation
rates on 100 Myr time scales, as for example adopted by \citet{Bose18},
are not relevant for the role of feedback driven halo expansion.

Observationally different star formation indicators probe different
time scales.  H$\alpha$ traces star formation within the past $\sim 5$
Myr, while far ultra-violet (FUV) photons trace longer time scales
$\sim 100-300$ Myr \citep{Calzetti13}.  Motivated by this as our
default time scales we consider 5 Myr and 200 Myr. On the spatial
scales of interest, the former is much smaller than the dynamical
time, while the latter is roughly equal to the dynamical time.  Recall
that in our simulations star formation is computed every 0.84 Myr (Age
of universe/$2^{14}$), so we can resolve star formation on a 5 Myr
timescale.

\subsection{Individual galaxies: test cases}

Fig.~\ref{fig:sfh} shows the star formation history (SFH, left), dark
matter circular velocity (middle), and gas fractions (right) for three
galaxies that have qualitatively different halo responses for $n=10$
(no change, expansion, contraction).  The red lines and points show
results for $n=10$, blue for $n=1$, and green for $n=0.1$.  The SFHs
are calculated from the ages of the star particles within the galaxy
at redshift $z=0$.  We see that the variation in the SFH is strongly
dependent on the star formation threshold, the time scale over which
we measure the star formation, and the mass of the galaxy. Higher
thresholds, shorter time scales, and lower masses result in more
bursty SFH. Below we will show that these trends hold for the full
sample of 20 haloes.

The middle panel shows the evolution of the dark matter circular
velocity at 1 per cent of the $z=0$ virial radius, normalized to the
virial velocity of the DMO simulation at $z=0$. Results for DMO
simulations (scaled by $\sqrt{1-\fbar}$) are shown with black
circles. The time evolution is typically quite smooth, both when the
halo expands and when it contracts. The upper panel shows a dwarf
galaxy (g1.18e10) that shows mild expansion for all three thresholds
and slightly more expansion for higher $n$. The middle panel shows a
galaxy (g1.59e11) that undergoes strong expansion for $n=10$, while
hardly any change for $n=1$ and $n=0.1$. The lower panel shows a Milky
Way mass galaxy (g7.08e11) which undergoes strong contraction for
$n=0.1$ and $n=1$, and mild contraction for $n=10$.  We thus see a
correlation between the burstiness of star formation on sub-dynamical
time scales and the halo response. More bursty SFHs yield lower
central dark matter densities, as expected from analytic models
\citep{Pontzen12,Dutton16b}

%% FIGURE 10
%------------------------------------------
\begin{figure*}
  \includegraphics[width=0.32\textwidth]{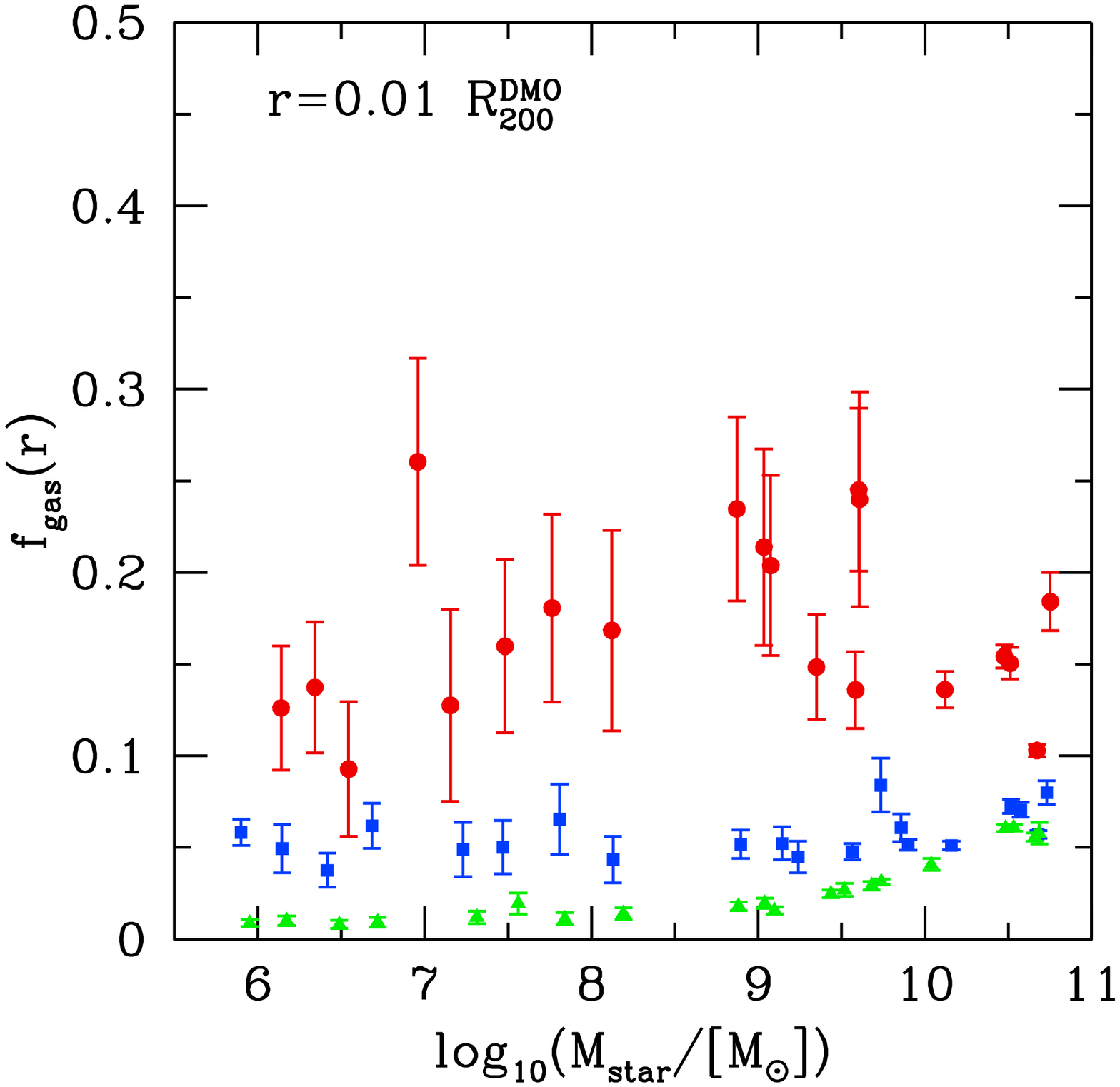}
  \includegraphics[width=0.32\textwidth]{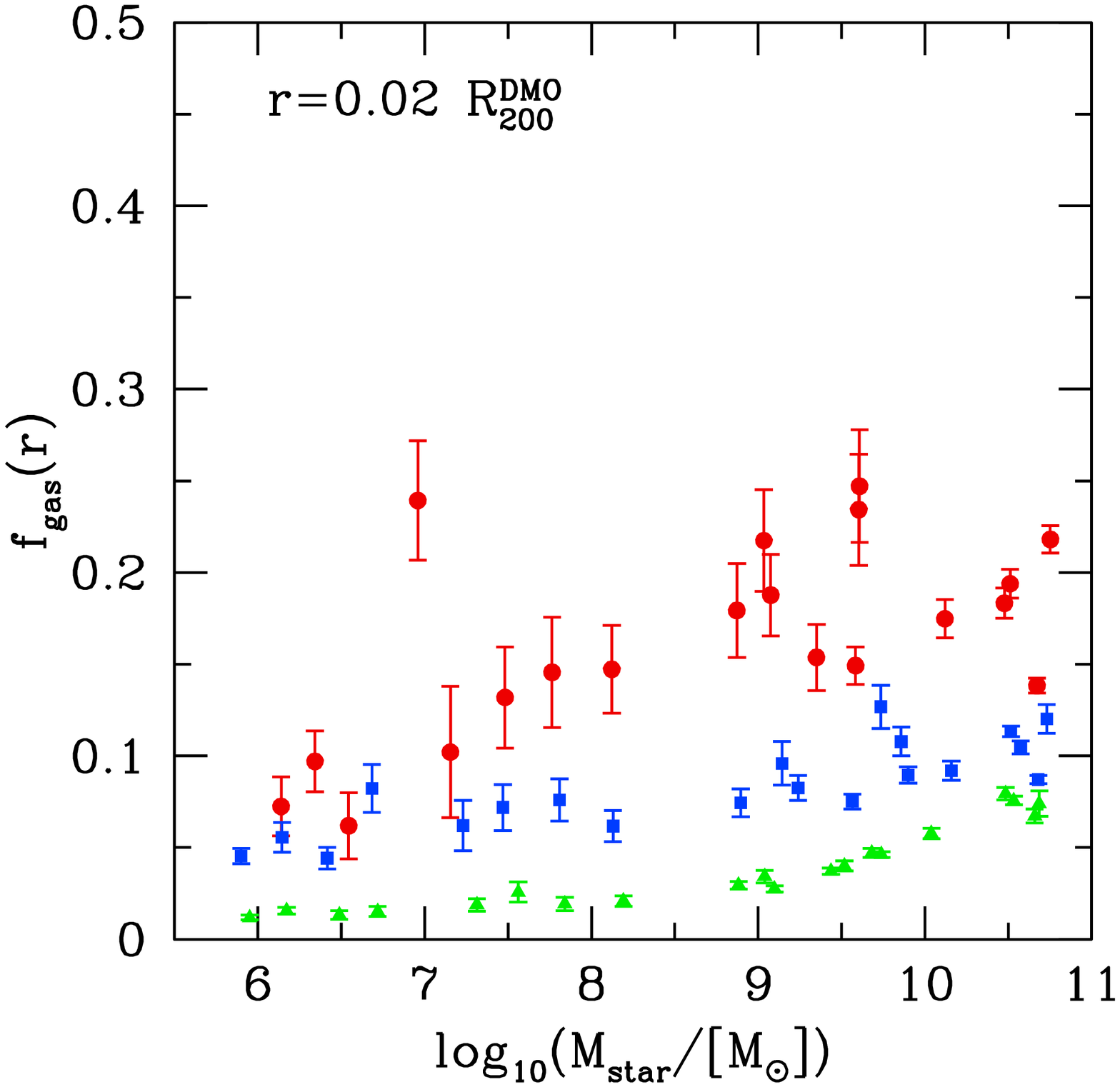}
  \includegraphics[width=0.32\textwidth]{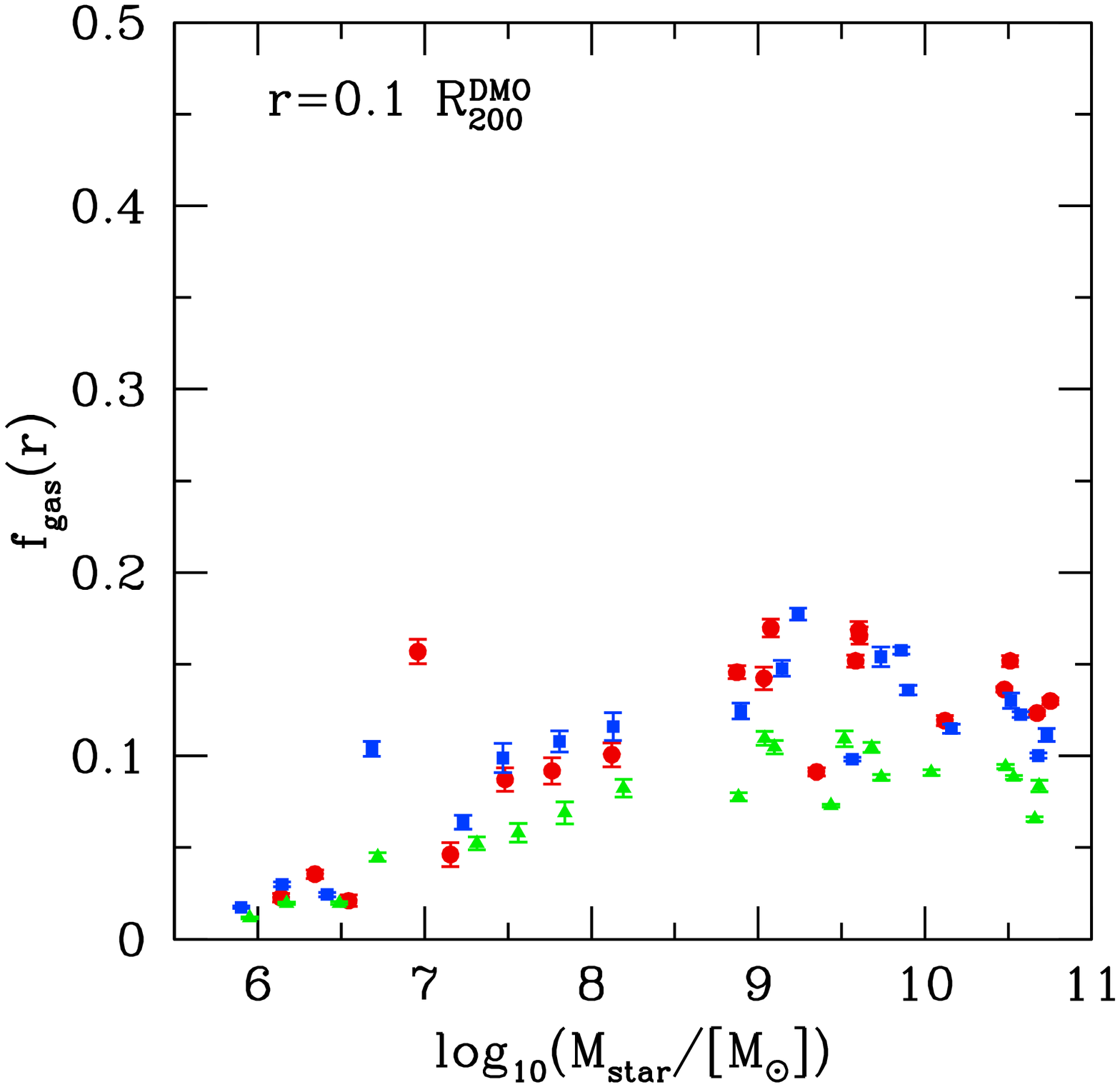}
  \includegraphics[width=0.32\textwidth]{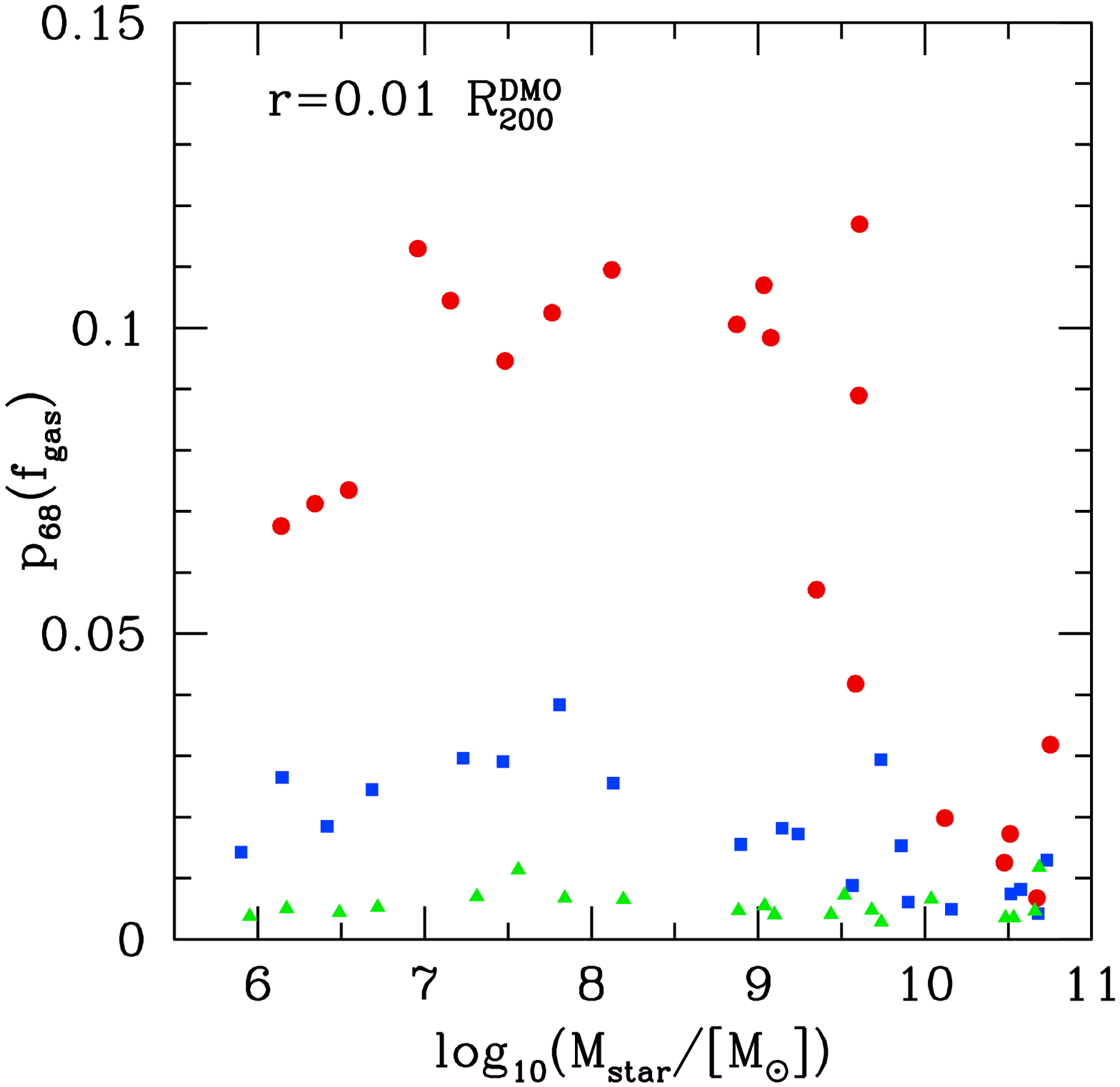}
  \includegraphics[width=0.32\textwidth]{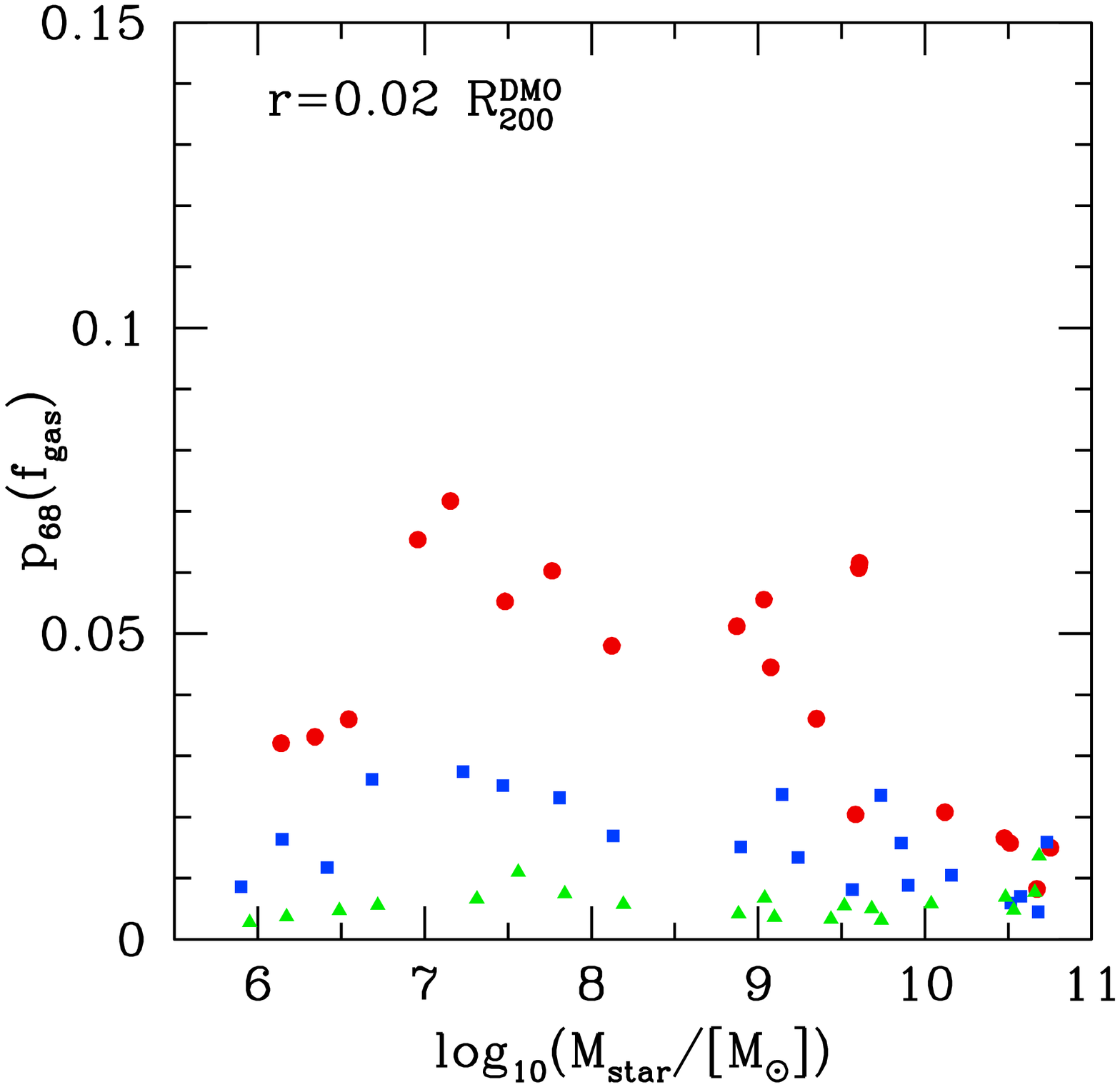}
  \includegraphics[width=0.32\textwidth]{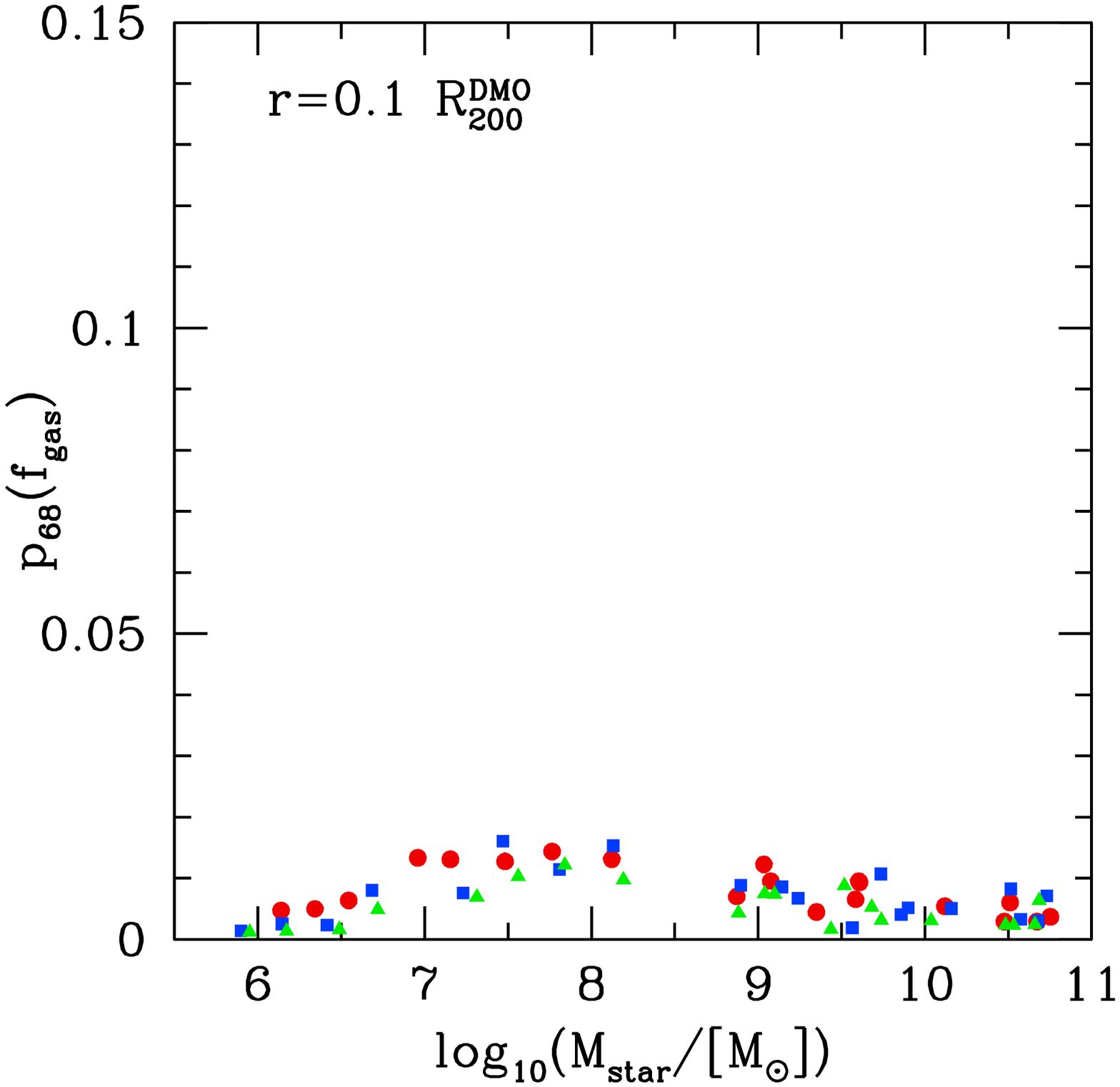}
  \caption{Mean gas fraction (upper panels) and variability in gas
    fraction (lower panels) versus $z=0$ stellar mass.  The colour and
    symbol type  refers to the star formation threshold: $n=10$ (red
    circles), $n=1$ (blue squares), $n=0.1$ (green triangles).  For
    each galaxy the gas fraction ($f_{\rm gas}(r)=M_{\rm gas}(r)/M_{\rm
      tot}(r)$) is measured within a fixed physical radius $r$ in the
    most massive progenitor. The mean gas fraction is averaged over
    time.  The error bars enclose 68 per cent of the variation in
    $f_{\rm gas}$ about the mean relation versus time.  For low threshold
    ($n=0.1$) the mean and variability in gas fraction are too small
    to have a meaningful impact on the dark halo structure. For $n=10$
    the mean gas fraction is $f_{\rm gas}\simeq 0.2$ and variability
    $\simeq 0.1$ in haloes that undergo the most expansion.}
  \label{fig:fgas}
\end{figure*}
%------------------------------------------

As discussed above, another key requirement for outflows to expand the
halo is for there to be sufficient gas at small radii.  The right
panels of Fig.~\ref{fig:sfh} show the gas fraction $(M_{\rm
  gas}/M_{\rm tot})$ measured at 1 per cent of the $z=0$ DMO virial
radius. This radius is indicated in the top left corner of the panels
and varies from $0.5$ kpc for g1.18e10 to $2.0$ kpc for g7.08e11.  We
see a clear trend for higher average gas fractions and higher
variability for higher star formation thresholds. This is expected
since for lower thresholds the gas turns into stars before it can get
very dense. We also see that the lowest mass galaxy (g1.18e10) has
lower gas fraction variability than the middle mass galaxy (g1.59e11),
which qualitatively explains the differences in expansion in these two
galaxies. For the Milky Way mass galaxy, the differences in halo
contraction start around a time of 3 Gyr. The $n=10$ simulation has
significant variations in the gas fraction around this time, which
likely prevents the strong contraction that occurs in the low
threshold simulations.

\subsection{Individual galaxies: full sample}
We now expand the results of the previous section to the full sample
of 20 haloes. Fig.~\ref{fig:sig_ssfr2} shows the standard deviation of
the SFH, $\sigma_{\rm SFH}$, in units of the mean SFH, $\mu_{\rm
  SFH}$.  This is a measure of how bursty the star formation is, and
can be used to quantify analytic SFHs used in interpreting galaxy
observations.  On a 5 Myr time scale (left panel), there is a roughly
constant shift in the scatter between the three thresholds, with more
bursty SFH  at all masses when the star formation threshold is higher.
On a 200 Myr time scale (right panel), there is not much difference in
the SFH for galaxies above a stellar mass of $\sim 10^8\,\Msun$.
Recall that a mass scale of $\Mstar\sim10^9\,\Msun$ is where there is
the most halo expansion. This confirms that variations in star
formation (and hence gas fractions) on a dynamical time scale do not
play a role in halo expansion. Variations in star formation needs to
occur on a sub-dynamical time scale in order to drive a halo response.
For both short and long time scales, we see that lower mass galaxies
have systematically more bursty SFHs.  Comparing this plot with the
halo response in Fig.~\ref{fig:alpha} we see that the burstiness of
star formation by itself does not determine the halo response, since
the lowest mass galaxies have the most bursty SFHs, yet they have no
change in the dark matter profile.

We have estimated the uncertainty on the SFH by using the Poisson
error on the number of star particles formed. This error is likely an
upper limit, because only a fraction of the gas eligible to form stars
actually turns into stars. The errors are only significant in the four
lowest mass galaxies ($\Mstar < 10^7\Msun$) and small timescales due
to there being only a few thousand star particles formed.  The
uncertainties are lower for higher $n$, because here the star
formation is concentrated into a few bursts, which are well resolved
(see top left panel in Fig.~\ref{fig:sfh}).  While for low $n$, the
star formation is roughly uniform in time, and there are only a
handful of star formation events in each 5 Myr time interval. We thus
conclude that the trend for more bursty star formation in lower mass
galaxies is not a manifestation of numerical discreteness.
   
%% FIGURE 11
%------------------------------------------
\begin{figure*}
  \includegraphics[width=0.45\textwidth]{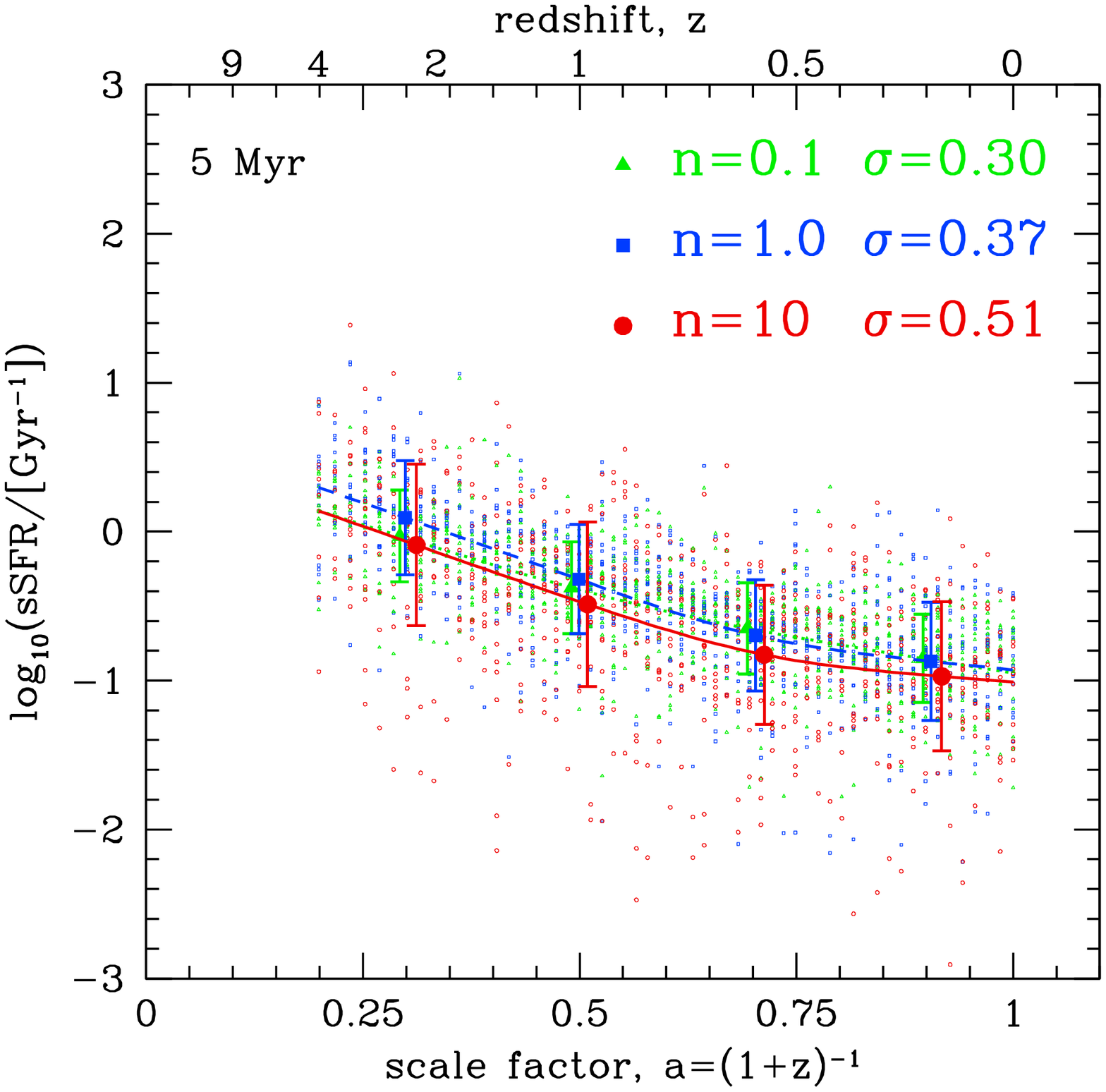}
  \includegraphics[width=0.45\textwidth]{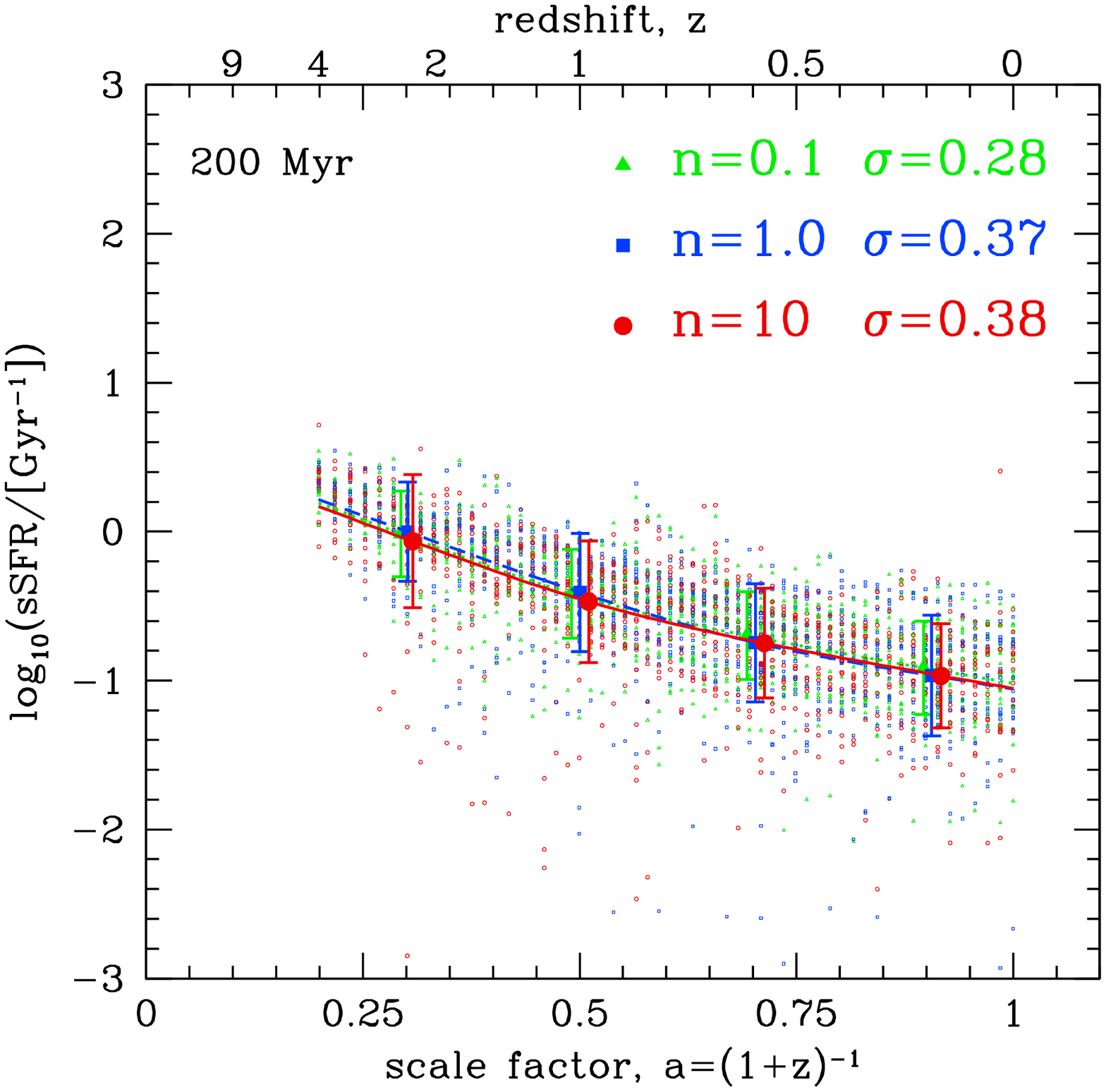}
  \caption{Specific star formation rate versus scale factor. Each
    small point is a separate galaxy. The large points show the mean
    sSFR in four scale factor bins. The lines show the interpolated
    relation with respect to which the standard deviation in
    $\log_{10}({\rm sSFR})$ is measured. The standard deviation for the
    different star formation thresholds is given in the top right
    corner. Star formation rates are measured over a 5 Myr time scale
    (left), and a 200 Myr time scale (right).}
  \label{fig:sig_ssfr3}
\end{figure*}
%------------------------------------------

%% FIGURE 12
%------------------------------------------
\begin{figure*}
  \includegraphics[width=0.45\textwidth]{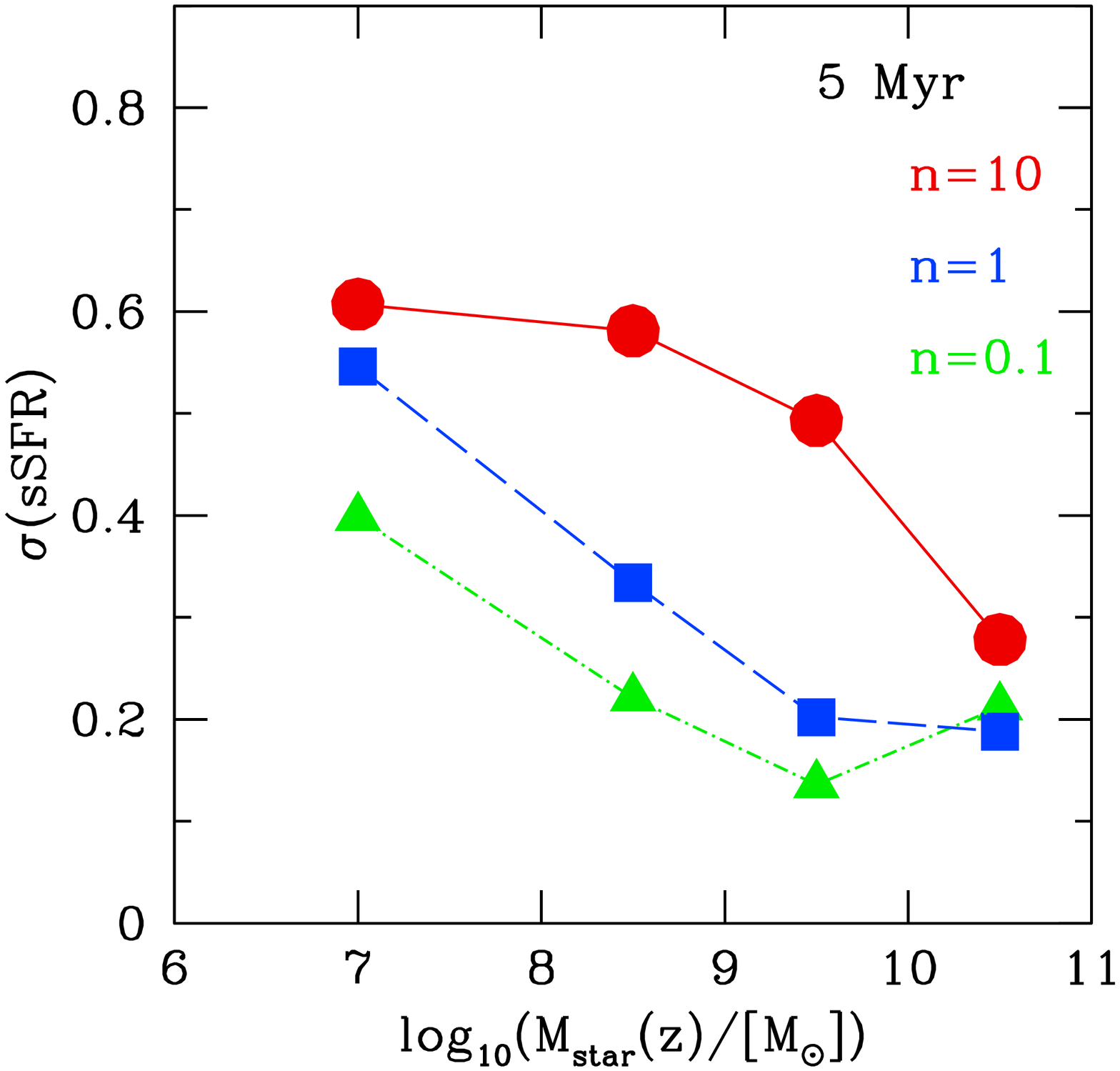}
  \includegraphics[width=0.45\textwidth]{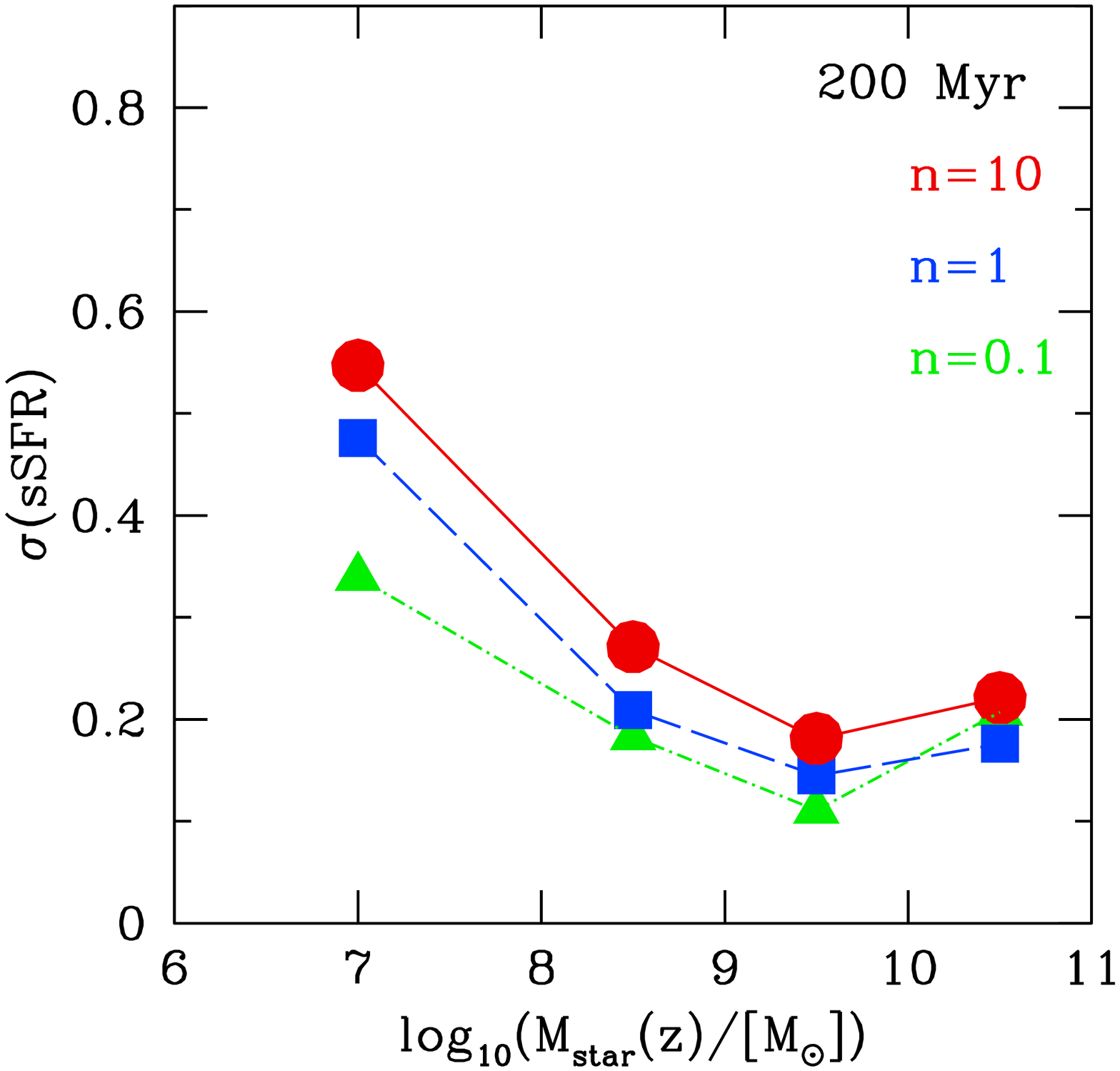}
  \caption{Scatter in specific star formation rate [dex] in bins of
    stellar mass for populations of galaxies at redshifts
    $z<4$. Colours and symbols refer to simulations with different
    star formation thresholds. Star formation rates are measured over
    a 5 Myr time scale (left), and a 200 Myr time scale (right).}
  \label{fig:sig_ssfr5}
\end{figure*}
%------------------------------------------

%% FIGURE 13
%------------------------------------------
\begin{figure*}
  \includegraphics[width=0.45\textwidth]{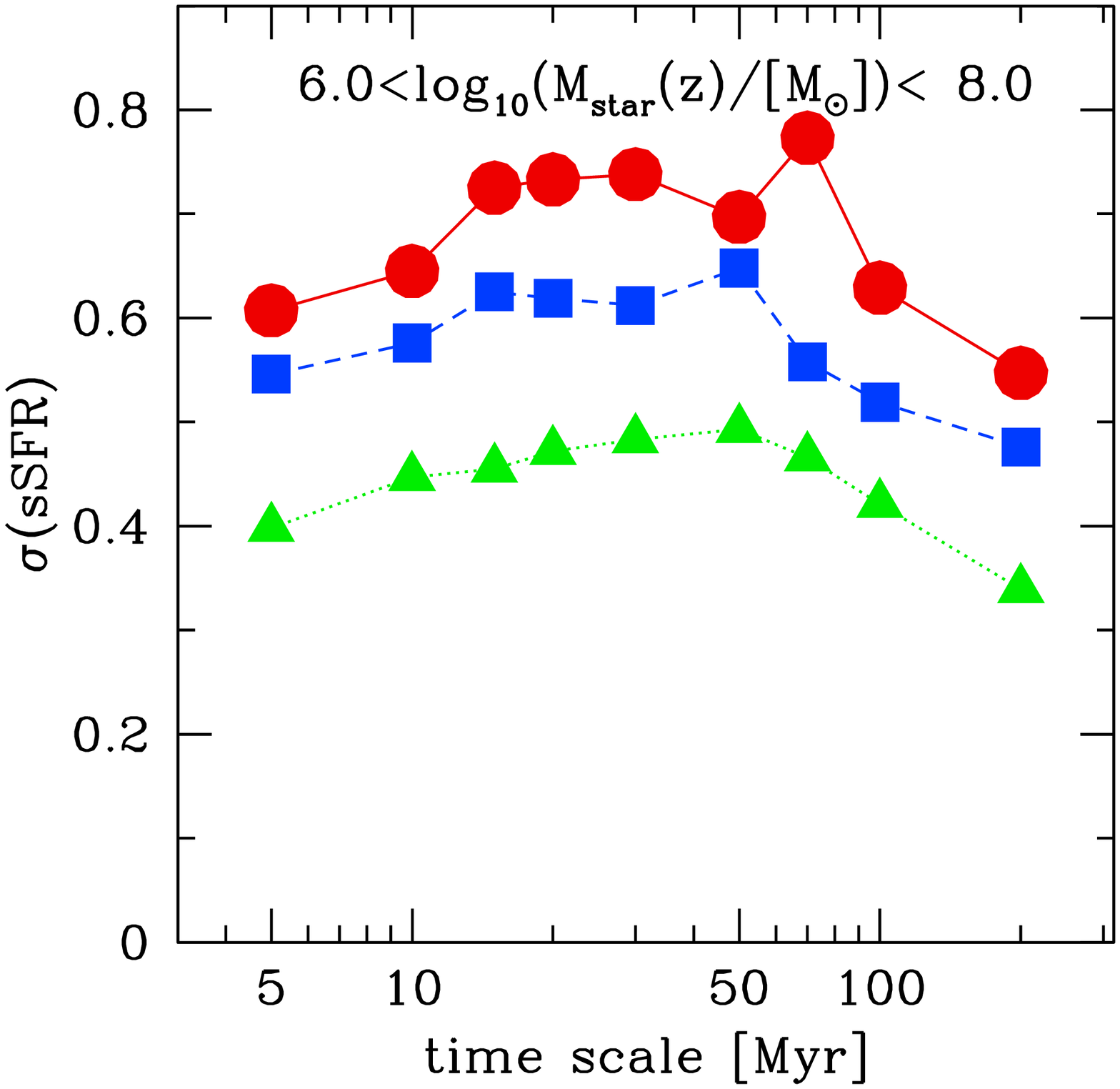}
  \includegraphics[width=0.45\textwidth]{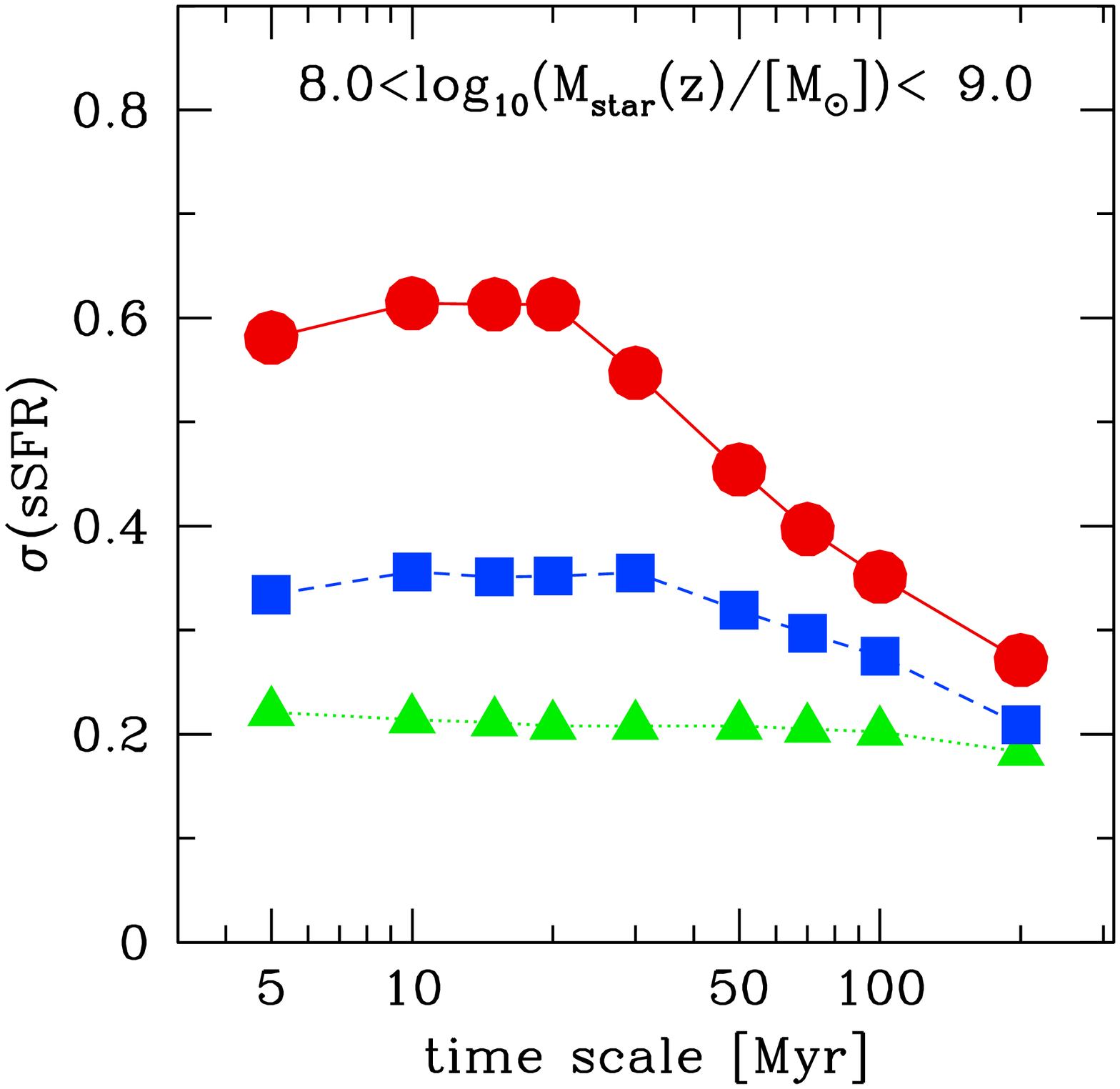}
  \includegraphics[width=0.45\textwidth]{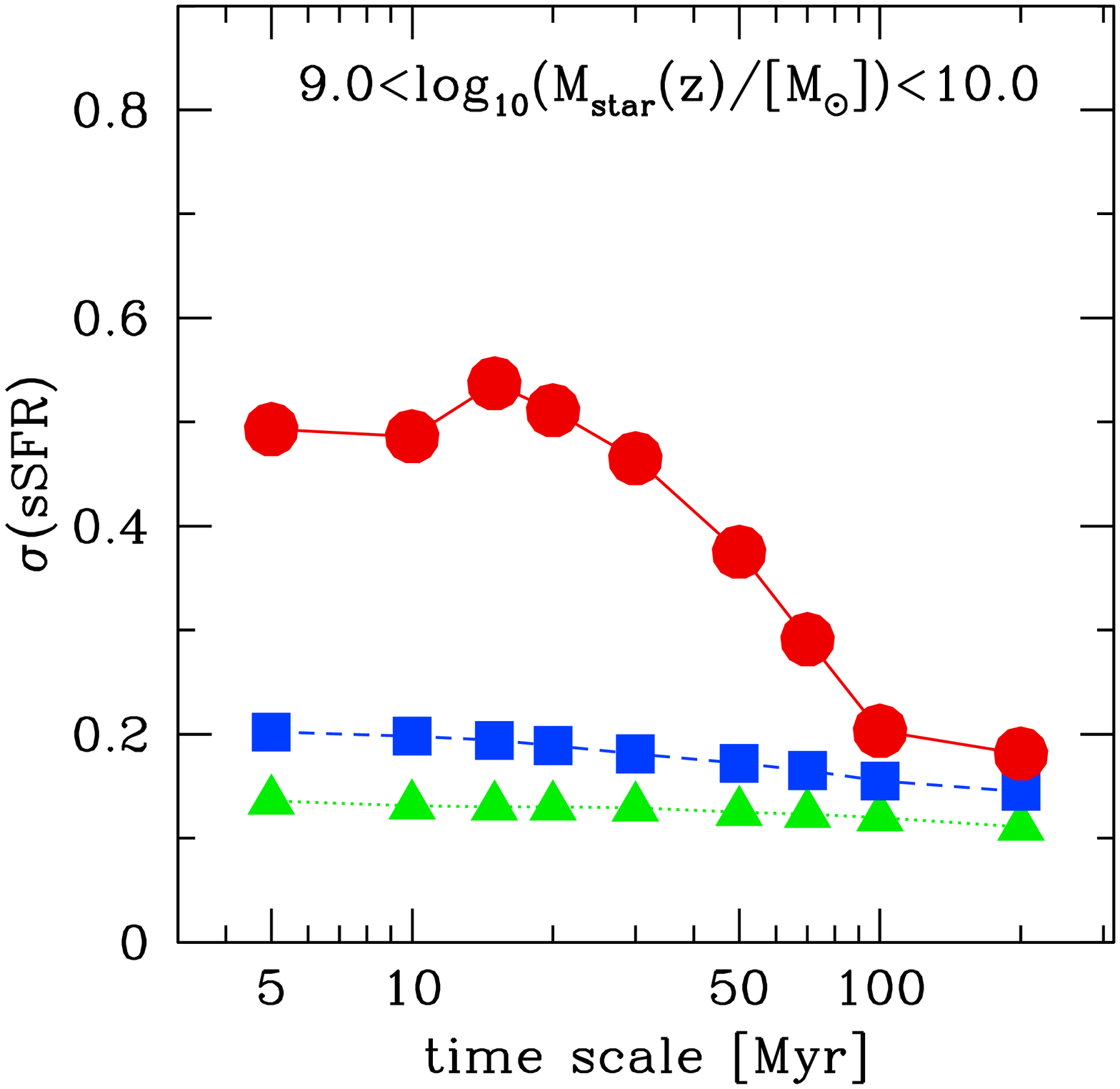}
  \includegraphics[width=0.45\textwidth]{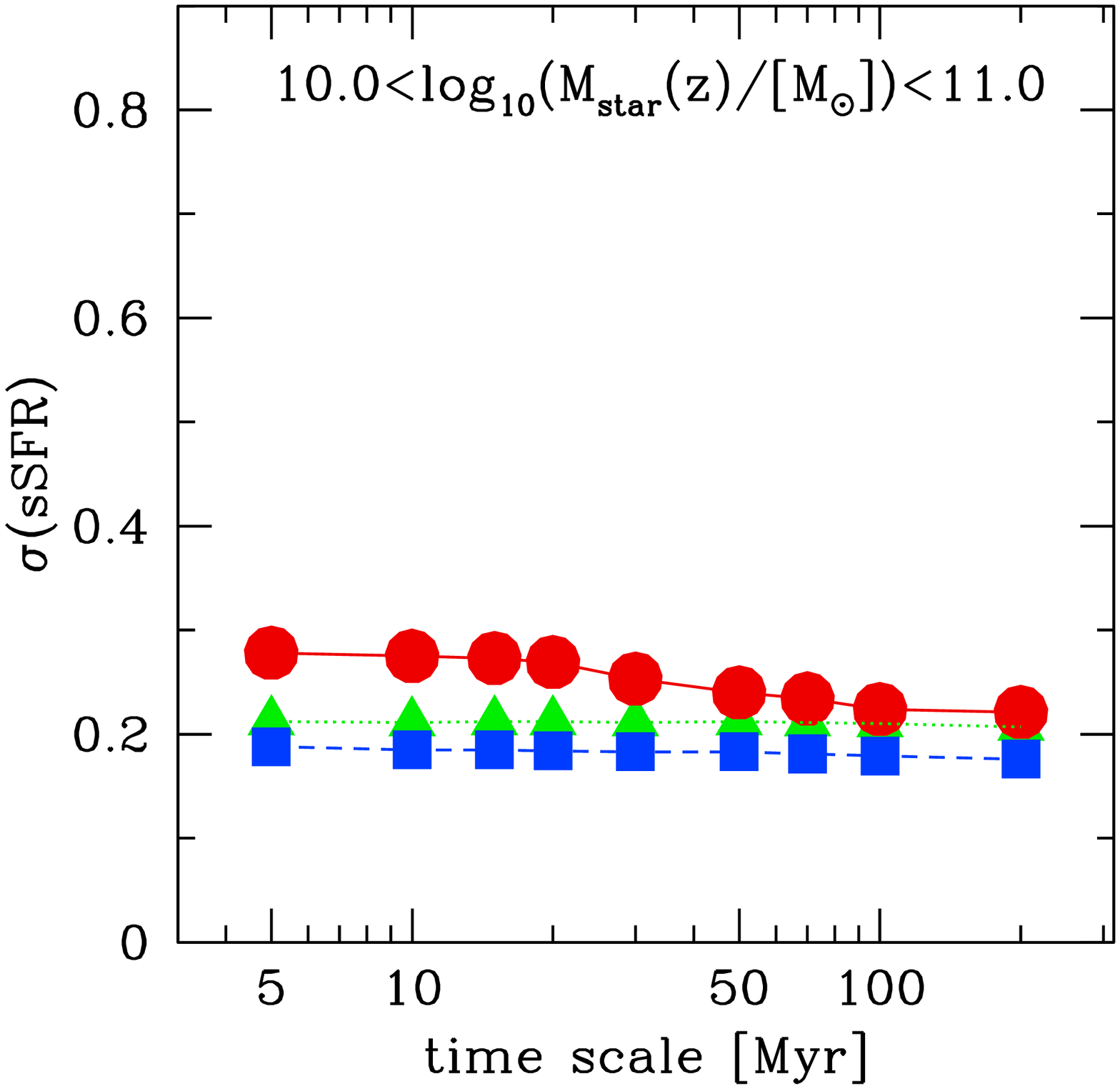}
  \caption{Scatter in specific star formation rate [dex] versus time
    scale over which star formation rate is measured. Each panel
    corresponds to a different range in stellar masses, $\Mstar(z)$,
    as indicated. Colours and symbols refer to simulations with
    different star formation thresholds: $n=10$ (red circles, solid
    lines); $n=1$ (blue squares, dashed lines); and $n=0.1$ (green
    triangles, dotted lines).}
  \label{fig:sig_ssfr4}
\end{figure*}
%------------------------------------------

The recent SFHs of individual galaxies can be
constrained using the ratio of H$\alpha$ to FUV fluxes \citep{Weisz12,
  Sparre17} or a combination of the 4000\AA\,  break, H$\delta_A$
indices and specific star formation rate SFR$/\Mstar$
\citep{Kauffmann14}.  The general conclusion from observations   is
that lower mass galaxies have more bursty recent SFHs.  This is
qualitatively consistent with our simulations with all $n$, and also
the FIRE simulations \citep{Sparre17}, which have a high $n$.

Fig.~\ref{fig:fgas} shows the mean (upper panels) and variability
(lower panels) of gas fractions (measured over cosmic time) versus the
stellar mass at $z=0$.  The gas fraction $f_{\rm gas}(r)=M_{\rm
  gas}(r)/M_{\rm tot}(r)$ is measured within a fixed physical
radius. Each point corresponds to the evolution of a single
galaxy. The radii are 1 per cent (left), 2 per cent (middle), and 10
per cent (right) of the $z=0$ DMO virial radius.  The variability is
defined as enclosing 68\% of the points about the mean.  This plot
explains the differences in halo response we see between different
$n$, as well as at different radii.  At large radii (10 per cent of
the virial radius), there is less than 2 per cent variability in
$f_{\rm gas}$ for all galaxy masses and star formation thresholds.
Recall, that the expansive effects of gas outflows go like the square
of the outflow fraction, so we expect no expansive effects at large
radii.  Thus on these scales the dark halo structure will be
determined by the adiabatic contraction due to inflows. At smaller
radii we see higher gas fractions and variability for $n=10$, but
lower gas fractions and variability for lower thresholds.  For $n=10$
the trend of variability with mass follows the trend of halo response
with mass seen in Fig.~\ref{fig:alpha}.  For $n=0.1$ the variability
is less than a per cent, which explains why we see no halo expansion
in low mass galaxies, in spite of there being some scatter in the
SFHs.

Note that the time scale at which we are measuring variation (in gas
fractions) is limited by the frequency $\simeq 215\,$Myr of the
simulation outputs. At small radii this is not a sub-dynamical
time. We have tested a few simulations with more outputs, from which
we see sudden drops in gas fractions occurring on $\sim 25\,$Myr time
scales. However, the scatter about the mean gas fraction is comparable
when we use 13 Myr between outputs, or our default of $215\,$Myr.
Note that the SFHs in Figs.~\ref{fig:sfh} and \ref{fig:sig_ssfr2} are
measured from the birth time of the star particles and are thus not
effected by the output frequency.

\subsection{Star formation rates for samples of galaxies}

In order to probe the variability in star formation on short time
scales $\sim 10$ Myr, one can resort to samples of galaxies.  For
samples of galaxies one can observationally measure the scatter in the
star formation rate versus stellar mass relation, or equivalently the
specific star formation rate, sSFR$=$SFR$/\Mstar$.
Fig.~\ref{fig:sig_ssfr3} shows the evolution of sSFR for the main
galaxy in each of the 20 simulations, for $z<4$, sSFR$ > 0.001$
Gyr$^{-1}$, and for $\Mstar(z) > 10^6 \,\Msun$.  The points are colour
coded by the star formation threshold.  The large points show means in
bins of scale factor. The lines are spline interpolations of the mean
with respect to which we measure the standard deviation,
$\sigma(\log_{10}{\rm sSFR})$. The error bars show the scatter in 4
redshift bins, showing minimal redshift dependence.  On 200 Myr time
scales (right panel), the scatter for $n=10$ and $n=1$ is the same
($\simeq 0.38$ dex) and 0.1 dex larger than for $n=0.1$. On a 5 Myr
time scale (left panel), there is a 0.2 dex difference in the scatter
for $n=10$ and $n=0.1$.

The mass dependence of the scatter is shown in
Fig.~\ref{fig:sig_ssfr5} for a time scale of 5 (left) and 200 Myr
(right). Similar to Fig.~\ref{fig:sig_ssfr2}, we see that there is
more variation in the sSFR in lower mass galaxies, shorter time
scales, and higher star formation thresholds. The differences with
respect to the star formation threshold are most pronounced at the
stellar masses ($10^8 \lta \Mstar \lta 10^{10}\,\Msun$), where
interestingly the most expansion occurs for $n=10$ simulations. There
is a large difference (0.4 dex) in scatter between $n=10$ and $n=0.1$
simulations on 5 Myr time scales, but just a 0.1 dex difference for a
200 Myr time scale.

In order to determine the characteristic time scale for variability in
star formation in Fig.~\ref{fig:sig_ssfr4}, we plot the scatter in
sSFR versus the time scale the star formation rate is measured over,
ranging from 5 Myr to 200 Myr, and in 4 bins of stellar mass (as
indicated).  For all time scales and stellar masses, $n=10$
simulations have larger scatter than both $n=1$ and $n=0.1$.  The
scatter peaks at a $\sim 20$ Myr time scale, which indicates the
characteristic time scale for episodes of star formation in the NIHAO
simulations.

%% SECTION 6
\section{Summary}

We use 20 sets of cosmological hydrodynamical simulations from the
NIHAO project \citep{Wang15} to investigate the impact of the star
formation threshold, $n$, on the response of the dark matter halo to
galaxy formation. We consider three values: $n=10[{\rm cm}^{-3}]$ (the
fiducial NIHAO), $n=1$, and $n=0.1$.  The masses of the central dark
matter haloes in our study cover the range $10^{10} \lta M_{200} \lta
10^{12} \,\Msun$ at redshift $z=0$.  We summarize our results as
follows:

\begin{itemize}
\item  For the $n=10$ and $n=1$ simulations the stellar to halo mass
  relations are consistent with halo abundance matching for redshifts
  $z<4$, without modification of any other parameter
  (Fig.~\ref{fig:msmh}).  For the $n=0.1$ simulations the feedback is
  too efficient, resulting in stellar masses that are too low. This is
  easily fixed by reducing the feedback efficiency from young stars
  from 0.13 to 0.04.

\item The dark matter density profiles at redshift $z=0$ show
  significant departures from dissipationless DMO simulations
  (Fig.~\ref{fig:rho}), with both contraction and expansion.

 \item The halo response at small radii ($\sim 1$ per cent of the
   virial radius) is more strongly correlated with $\Mstar/M_{200}$
   than either stellar mass or halo mass alone for all $n$
   (Fig.~\ref{fig:alpha}). 

\item Simulations with $n=0.1$ do not result in halo expansion (beyond
  adiabatic mass loss) at any mass scale we study
  (Fig.~\ref{fig:rho}). Simulations with $n=1$ largely follow the halo
  response of $n=0.1$, but with a small amount of expansion at $0.001
  \lta \Mstar/M_{200} \lta 0.01$. Simulations with $n=10$ result in
  strong expansion for $0.001 \lta \Mstar/M_{200} \lta 0.01$ and
  weaker contraction (than $n=0.1$) for  $\Mstar/M_{200} \gta 0.3$
  (Fig.~\ref{fig:alpha}). 

\item Field galaxies in the local group with stellar masses $10^6 \lta
  \Mstar \lta 10^8 \Msun$ have circular velocities at the half-light
  radius consistent with $n=10$ simulations, but a factor of $\simeq
  1.4$ lower than $n=0.1$ simulations (Fig.~\ref{fig:tbtf}).

\item Field galaxies with stellar masses $10^9 \lta \Mstar \lta
  10^{10} \Msun$ have dark matter circular velocities at 2 kpc a
  factor of $1.5$ lower than predicted by $n=1$ and $n=0.1$
  simulations but a factor of $1.15$ higher than $n=10$ simulations 
  (Fig.~\ref{fig:vr_sparc}).
  
\item For Milky Way mass galaxies, the simulations are  consistent
  with the observed circular velocity at 8 kpc and 60 kpc.  All haloes
  contract, with more contraction at smaller radii.  Simulations with
  $n=0.1$ and $n=1$ have almost identical halo contraction, while
  simulations with $n=10$ have a factor $\sim 2$ less contraction at
  small scales (Fig.~\ref{fig:mw}).

\item Lower $n$ simulations have lower mean, and lower variability in
  the  gas fractions within small radii (Figs.~\ref{fig:sfh} \&
  \ref{fig:fgas}). For the galaxies that experience the most halo
  expansion in the $n=10$ simulations the variations in gas fractions
  are $\simeq 0.1$. Each such outflow and inflow cycle is expected to
  expand the dark matter orbits by roughly a per cent, so many such
  cycles are required to significantly expand the dark matter
  halo. For the $n=0.1$ simulations the variability in gas fractions
  is $\sim 0.01$, and so the expansive effect of each cycle is $\sim
  10^{-4}$, and thus negligible.  The average gas fractions and the
  variability in the gas fractions (for populations of galaxies) could
  be measured with radio observations.

\item Star formation is more bursty in lower mass galaxies, and for
  simulations with higher $n$ (Figs.~\ref{fig:sig_ssfr2} \&
  \ref{fig:sig_ssfr5}).  

\item For star formation feedback to drive halo expansion the
  variability must occur on a sub-dynamical time scale (i.e.,  much
  less than 50 Myr at one percent of the virial radius). Measuring
  star formation rates on a 100 Myr time scale, as done by
  \citet{Bose18}, is thus not relevant to the problem of feedback driven
  halo expansion.
   
\item For $n=0.1$ simulations the scatter in sSFR is only weakly
  dependent on the time scale over which the SFR is
  measured. However, for $n=10$ simulations, shorter time scales
  result in larger scatter down to a characteristic time scale of
  $\sim 20$ Myr (Fig.~\ref{fig:sig_ssfr4}).

\item Bursty star formation on a short time scale is a necessary but not
  sufficient condition for halo expansion to occur.  This is readily
  apparent when one sees that the lowest mass galaxies we simulate
  $\Mstar\sim 10^6\Msun$ have the most bursty star formation
  (Fig.~\ref{fig:sig_ssfr2}), yet the halo response is weak
  (Fig.~\ref{fig:alpha}).

\item At the stellar mass range that undergoes the most halo expansion
  $10^8 \lta \Mstar \lta 10^9 \,\Msun$, there is 0.4 dex more scatter
  in the sSFR (when measured on $\lta 20$ Myr time scales) for
  $n=10$ simulations compared to $n=0.1$ simulations
  (Fig.~\ref{fig:sig_ssfr4}).  In principle star formation rates on
  these time scales could be measured with H$\alpha$ emission,
  provided that extinction does not impact the scatter in sSFR too
  severely.
 
\end{itemize}

Our study reconciles the conflicting results in the literature from
cosmological simulations for how the dark matter halo responds to
galaxy formation. In addition to the ratio between stellar mass and
halo mass \citep[e.g.,][]{DiCintio14}, we have shown that the halo
response is a strong function of the star formation threshold.
Simulations that use low star formation thresholds ($n\lta 0.1$), such
as EAGLE \citep{Schaye15} find no significant change in haloes of mass
$10^{10}$ and $10^{11}\Msun$, and mild contraction in haloes of mass
$10^{12}\Msun$ \citep{Schaller15, Bose18}. This is exactly the same as
we find for our $n=0.1$ simulations.  Simulations with high star
formation thresholds ($n\gta 10$) such as \citet{Governato10,
  DiCintio14, Tollet16, Chan15} result in halo expansion in dwarf
galaxies (halo masses $\sim 10^{10}$ to $\sim 10^{11}\Msun$) and
contraction in Milky Way mass haloes ($\sim 10^{12}\Msun$). 

A similar conclusion has been reached independently by
\citet{Benitez18} using simulations with the EAGLE code.  These
authors take the dependence of halo structure on the star formation
threshold with a pessimistic view of being able to predict the
structure of CDM haloes.  We are more optimistic, because we have
shown that there are large differences in the gas fractions and star
formation rates in simulations with different star formation
thresholds.   Thus it should be possible in the near future to
observationally distinguish between different star formation
thresholds, and more generally to calibrate the free parameters of the
sub-grid model for star formation and feedback.

Indeed, in \citet{Buck19} we use our simulations to show that the
spatial clustering strength of young stars depends on the star
formation threshold. The observed clustering from the HST Legacy
Extragalactic UV Survey \citep[LEGUS][]{Grasha17} is inconsistent with
a low threshold ($n<1$ [cm$^{-3}$]) and strongly favours a high
threshold ($n>10$ [cm$^{-3}$]).

  \section*{Acknowledgements}
We thank the referee whose report helped to improve the paper.
This research was carried out on the High Performance Computing
resources at New York University Abu Dhabi; on the  {\sc theo} cluster
of the Max-Planck-Institut f\"ur Astronomie and on the {\sc hydra}
clusters at the Rechenzentrum in Garching.
The authors gratefully acknowledge the Gauss Centre for Supercomputing
e.V. (www.gauss-centre.eu) for funding this project by providing
computing time on the GCS Supercomputer SuperMUC at Leibniz
Supercomputing Centre (www.lrz.de).
TB acknowledges support from the Sonderforschungsbereich SFB 881 “The
Milky Way System” (subproject A2) of the German Research Foundation
(DFG).
AO is funded by the Deutsche Forschungsgemeinschaft (DFG, German  
Research Foundation) -- MO 2979/1-1.

%%%%%%%%%%%%%%%%%%%%%%%%%%%%%%%%%%%%%%%%%%%%%%%%%%%%%%%%%%%%%%%%%%%%%%
%%  REFERENCES
%%%%%%%%%%%%%%%%%%%%%%%%%%%%%%%%%%%%%%%%%%%%%%%%%%%%%%%%%%%%%%%%%%%%%% 
%\vspace{-0.5cm}


\begin{thebibliography}{99}

%The Average Star Formation Histories of Galaxies in Dark Matter Halos
% from z = 0-8
\bibitem[Behroozi et al.(2013)]{Behroozi13} Behroozi, P.~S.,
  Wechsler, R.~H., \& Conroy, C.\ 2013, \apj, 770, 57
  
% Baryon-induced dark matter cores in the EAGLE simulations
\bibitem[Benitez-Llambay et al.(2018)]{Benitez18} Benitez-Llambay, A., Frenk, C.~S., Ludlow, A.~D., \& Navarro, J.~F.\ 2018, arXiv:1810.04186 
  
%Contraction of dark matter galactic halos due to baryonic infall
\bibitem[Blumenthal et al.(1986)]{Blumenthal86} Blumenthal,
  G.~R., Faber, S.~M., Flores, R., \& Primack, J.~R., 1986, ApJ, 301, 27

%No cores in dark matter-dominated dwarf galaxies with bursty star formation histories  
\bibitem[Bose et al.(2018)]{Bose18} Bose, S., Frenk, C.~S., Jenkins, A., et al.\ 2018, arXiv:1810.03635 

%A Direct Dynamical Measurement of the Milky Way's Disk Surface Density Profile, Disk Scale Length, and Dark Matter Profile at 4 kpc <~ R <~ 9 kpc
\bibitem[Bovy \& Rix(2013)]{Bovy13} Bovy, J., \& Rix, H.-W.\ 2013, \apj, 779, 115 

% NIHAO XIII: Clumpy discs or clumpy light in high-redshift galaxies?
\bibitem[Buck et al.(2017)]{Buck17} Buck, T., Macci{\`o}, A.~V., Obreja, A., et al.\ 2017, \mnras, 468, 3628 

%NIHAO XV: The environmental impact of the host galaxy on galactic satellite and field dwarf galaxies
\bibitem[Buck et al.(2018)]{Buck18} Buck, T., Macci{\`o}, A.~V., Dutton, A.~A., Obreja, A., \& Frings, J.\ 2018, MNRAS in press, arXiv:1804.04667 


\bibitem[Buck et al.(2018)]{Buck19} Buck, T., Dutton, A.~A., \& Macci{\`o}, A.~V.\ 2018, arXiv:1812.05613 
  
%Small-Scale Challenges to the ΛCDM Paradigm  
\bibitem[Bullock \& Boylan-Kolchin(2017)]{Bullock17} Bullock, J.~S., \& Boylan-Kolchin, M.\ 2017, \araa, 55, 343 

%Star Formation Rate Indicators
\bibitem[Calzetti(2013)]{Calzetti13} Calzetti, D.\ 2013, Secular Evolution of Galaxies, 419 
  
%The impact of baryonic physics on the structure of dark matter haloes: the view from the FIRE cosmological simulations
 \bibitem[Chan et al.(2015)]{Chan15} Chan, T.~K., Kere{\v s}, 
D., O{\~n}orbe, J., et al.\ 2015, \mnras, 454, 2981 

%The EAGLE simulations of galaxy formation: calibration of subgrid physics and model variations
\bibitem[Crain et al.(2015)]{Crain15} Crain, R.~A., Schaye, J., Bower, R.~G., et al.\ 2015, \mnras, 450, 1937 

%Simulating galactic outflows with thermal supernova feedback
\bibitem[Dalla Vecchia \& Schaye(2012)]{DallaVecchia12} Dalla Vecchia, C., \& Schaye, J.\ 2012, \mnras, 426, 140 
  
 % The dependence of dark matter profiles on the stellar-to-halo mass ratio: a prediction for cusps versus cores
  \bibitem[Di Cintio et al.(2014a)]{DiCintio14} Di Cintio, A., Brook, 
C.~B., Macci{\`o}, A.~V., et al.\ 2014, \mnras, 437, 415 

%Dark halo response and the stellar initial mass function in early-type and late-type galaxies
\bibitem[Dutton et al.(2011)]{Dutton11} Dutton, A.~A., Conroy, C., van den Bosch, F.~C., et al.\ 2011, \mnras, 416, 322 

%Cold dark matter haloes in the Planck era: evolution of structural parameters for Einasto and NFW profiles
\bibitem[Dutton \& Macci{\`o}(2014)]{Dutton14} Dutton,
  A.~A., \& Macci{\`o}, A.~V.\ 2014, \mnras, 441, 3359

% NIHAO V: too big does not fail - reconciling the conflict between ΛCDM predictions and the circular velocities of nearby field galaxies
\bibitem[Dutton et al.(2016a)]{Dutton16a} Dutton, A.~A., Macci{\`o}, A.~V., Frings, J., et al.\ 2016a, \mnras, 457, L74
    
% NIHAO IX: The role of gas inflows and outflows in driving the contraction and expansion of cold dark matter haloes
\bibitem[Dutton et al.(2016b)]{Dutton16b} Dutton, A.~A., Macci{\`o}, A.~V., Dekel, A., et al.\ 2016b, \mnras, 461, 2658 

%NIHAO XII: galactic uniformity in a ΛCDM universe
\bibitem[Dutton et al.(2017)]{Dutton17} Dutton, A.~A., Obreja, A., Wang, L., et al.\ 2017, \mnras, 467, 4937 

%NIHAO XVII: The diversity of dwarf galaxy kinematics and implications for the HI velocity function
\bibitem[Dutton et al.(2019)]{Dutton19} Dutton, A.~A., Obreja, A., \& Macci{\`o}, A.~V.\ 2019, \mnras, 482, 5606 
    
%Dark Halos: The Flattening of the Density Cusp by Dynamical Friction
\bibitem[El-Zant et al.(2001)]{El-Zant01} El-Zant, A., Shlosman, I., \& Hoffman, Y.\ 2001, \apj, 560, 636 
  
 % Too big to fail in the Local Group
\bibitem[Garrison-Kimmel et al.(2014)]{Garrison-Kimmel14} Garrison-Kimmel, S., Boylan-Kolchin, M., Bullock, J.~S., \& Kirby, E.~N.\ 2014, \mnras, 444, 222 

%	The evolution of substructure - I. A new identification method
\bibitem[Gill et al.(2004)]{Gill04} Gill, S.~P.~D.,  Knebe, A., \& Gibson, B.~K.\ 2004, \mnras, 351, 399
  
%Response of Dark Matter Halos to Condensation of Baryons: 
%Cosmological Simulations and Improved Adiabatic Contraction Model
\bibitem[Gnedin  et   al.(2004)]{Gnedin04}  Gnedin,  O.~Y.,
Kravtsov, A.~V., Klypin, A.~A., \& Nagai, D.\ 2004, \apj, 616, 16

%Bulgeless dwarf galaxies and dark matter cores from supernova-driven outflows
\bibitem[Governato et al.(2010)]{Governato10} Governato, F., Brook, C., Mayer, L., et al.\ 2010, \nat, 463, 203 

%Cuspy no more: how outflows affect the central dark matter and baryon distribution in Λ cold dark matter galaxies
%\bibitem[Governato et al.(2012)]{Governato12} Governato, F., 
%  Zolotov, A., Pontzen, A., et al.\ 2012, \mnras, 422, 1231

%The Auriga Project: the properties and formation mechanisms of disc galaxies across cosmic time
\bibitem[Grand et al.(2017)]{Grand17} Grand, R.~J.~J., G{\'o}mez, F.~A., Marinacci, F., et al.\ 2017, \mnras, 467, 179 

%The Hierarchical Distribution of the Young Stellar Clusters in Six Local Star-forming Galaxies    
\bibitem[Grasha et al.(2017)]{Grasha17} Grasha, K., Calzetti, D., Adamo, A., et al.\ 2017, \apj, 840, 113 
  
 %NIHAO VIII: Circum-galactic medium and outflows - The puzzles of HI and OVI gas distributions
\bibitem[Gutcke et al.(2017)]{Gutcke17} Gutcke, T.~A., Stinson, G.~S., Macci{\`o}, A.~V., Wang, L., \& Dutton, A.~A.\ 2017, \mnras, 464, 2796 
  
%Galaxies on FIRE (Feedback In Realistic Environments): stellar feedback explains cosmologically inefficient star formation
\bibitem[Hopkins et al.(2014)]{Hopkins14} Hopkins, P.~F., Kere{\v 
s}, D., O{\~n}orbe, J., et al.\ 2014, \mnras, 445, 581 

%FIRE-2 simulations: physics versus numerics in galaxy formation
\bibitem[Hopkins et al.(2018)]{Hopkins18} Hopkins, P.~F., Wetzel, A., Kere{\v s}, D., et al.\ 2018, \mnras, 480, 800 
    
%  Quantitative constraints on starburst cycles in galaxies with stellar masses in the range 108-1010 M⊙
\bibitem[Kauffmann(2014)]{Kauffmann14} Kauffmann, G.\ 2014, \mnras, 441, 2717 

%The dynamics of isolated Local Group galaxies
\bibitem[Kirby et al.(2014)]{Kirby14} Kirby, E.~N., Bullock,  J.~S., Boylan-Kolchin, M., Kaplinghat, M.,  \& Cohen, J.~G.\ 2014, \mnras, 439, 1015 
  
%AHF: Amiga's Halo Finder
\bibitem[Knollmann \& Knebe(2009)]{Knollmann09} Knollmann, S.~R., \& Knebe, A.\ 2009, \apjs, 182, 608
  
%SPARC: Mass Models for 175 Disk Galaxies with Spitzer Photometry and Accurate Rotation Curves  
\bibitem[Lelli et al.(2016)]{Lelli16} Lelli, F., McGaugh, S.~S., \& Schombert, J.~M.\ 2016, \aj, 152, 157 

%Halo Expansion in Cosmological Hydro Simulations: Toward a Baryonic Solution of the Cusp/Core Problem in Massive Spirals
\bibitem[Macci{\`o} et al.(2012)]{Maccio12} Macci{\`o}, A.~V., Stinson, G., Brook, C.~B., et al.\ 2012, \apjl, 744, L9 

%NIHAO X: reconciling the local galaxy velocity function with cold dark matter via mock H I observations
\bibitem[Macci{\`o} et al.(2016)]{Maccio16} Macci{\`o}, A.~V., Udrescu, S.~M., Dutton, A.~A., et al.\ 2016, \mnras, 463, L69 

% The formation of disc galaxies in high-resolution moving-mesh cosmological simulations
\bibitem[Marinacci et al.(2014)]{Marinacci14} Marinacci, F., Pakmor, R., \& Springel, V.\ 2014, \mnras, 437, 1750

% Galactic star formation and accretion histories from matching galaxies to dark matter haloes
\bibitem[Moster et al.(2013)]{Moster13} Moster, B.~P., Naab, T., 
\& White, S.~D.~M.\ 2013, \mnras, 428, 3121 
  
  %EMERGE - an empirical model for the formation of galaxies since z ˜ 10
\bibitem[Moster et al.(2018)]{Moster18} Moster, B.~P., Naab, T., \& White, S.~D.~M.\ 2018, \mnras, 477, 1822 

%The unexpected diversity of dwarf galaxy rotation curves
\bibitem[Oman et al.(2015)]{Oman15} Oman, K.~A., Navarro, J.~F., Fattahi, A., et al.\ 2015, \mnras, 452, 3650 

%Non-circular motions and the diversity of dwarf galaxy rotation curves
\bibitem[Oman et al.(2019)]{Oman19} Oman, K.~A., Marasco, A., Navarro, J.~F., et al.\ 2019, \mnras, 482, 821 
  
%Planck 2013 results. XVI. Cosmological parameters
\bibitem[Planck Collaboration et al.(2014)]{Planck14} Planck Collaboration, Ade, P.~A.~R., Aghanim, N., et al.\ 2014, \aap, 571, A16

%How supernova feedback turns dark matter cusps into cores
\bibitem[Pontzen \& Governato(2012)]{Pontzen12} Pontzen, A., \& Governato, F.\ 2012, \mnras, 421, 3464

%The inner structure of ΛCDM haloes - I. A numerical convergence study  
\bibitem[Power et al.(2003)]{Power03} Power, C., Navarro, J.~F., Jenkins, A., et al.\ 2003, \mnras, 338, 14 
  
%The impact of local stellar radiation on the H I column density distribution
\bibitem[Rahmati et al.(2013)]{Rahmati13} Rahmati, A., Schaye, J., Pawlik, A.~H., \& Rai{\v c}evi$\grave c$, M.\ 2013, \mnras, 431, 2261 

%Mass loss from dwarf spheroidal galaxies: the origins of shallow dark
%matter cores and exponential surface brightness profiles
\bibitem[Read \& Gilmore(2005)]{Read05} Read, J.~I., \& Gilmore, G.\ 2005, \mnras, 356, 107
  
%Dark matter cores all the way down  
\bibitem[Read et al.(2016)]{Read16} Read, J.~I., Agertz, O., \& Collins, M.~L.~M.\ 2016, \mnras, 459, 2573 

%NIHAO - XIV. Reproducing the observed diversity of dwarf galaxy rotation curve shapes in ΛCDM
\bibitem[Santos-Santos et al.(2018)]{Santos-Santos18} Santos-Santos, I.~M., Di Cintio, A., Brook, C.~B., et al.\ 2018, \mnras, 473, 4392 
  
%The APOSTLE simulations: solutions to the Local Group's cosmic puzzles
\bibitem[Sawala et al.(2016)]{Sawala16} Sawala, T., Frenk, C.~S., Fattahi, A., et al.\ 2016, \mnras, 457, 1931 

%Baryon effects on the internal structure of ΛCDM haloes in the EAGLE simulations
\bibitem[Schaller et al.(2015)]{Schaller15} Schaller, M., Frenk, C.~S., Bower, R.~G., et al.\ 2015, \mnras, 451, 1247 

% The EAGLE project: simulating the evolution and assembly of galaxies and their environments
\bibitem[Schaye et al.(2015)]{Schaye15} Schaye, J., Crain,  R.~A., Bower, R.~G., et al.\ 2015, \mnras, 446, 521
  
%The Physical Origin of Long Gas Depletion Times in Galaxies
\bibitem[Semenov et al.(2017)]{Semenov17} Semenov, V.~A., Kravtsov, A.~V., \& Gnedin, N.~Y.\ 2017, \apj, 845, 133 

%A Catalog of Bulge+disk Decompositions and Updated Photometry for 1.12 Million Galaxies in the Sloan Digital Sky Survey
\bibitem[Simard et al.(2011)]{Simard11} Simard, L., Mendel, J.~T., Patton, D.~R., Ellison, S.~L., \& McConnachie, A.~W.\ 2011, \apjs, 196, 11 

%(Star)bursts of FIRE: observational signatures of bursty star formation in galaxies
\bibitem[Sparre et al.(2017)]{Sparre17} Sparre, M., Hayward, C.~C., Feldmann, R., et al.\ 2017, \mnras, 466, 88 
  
%Quantifying the heart of darkness with GHALO - a multibillion particle simulation of a galactic halo  
\bibitem[Stadel et al.(2009)]{Stadel09} Stadel, J., Potter, D., Moore, B., et al.\ 2009, \mnras, 398, L21 

  % Star formation and feedback in smoothed particle hydrodynamic
% simulations - I. Isolated galaxies
\bibitem[Stinson et al.(2006)]{Stinson06} Stinson, G., Seth, A., 
Katz, N., et al.\ 2006, \mnras, 373, 1074 

%Making Galaxies In a Cosmological Context: the need for early stellar feedback
\bibitem[Stinson et al.(2013)]{Stinson13} Stinson, G.~S., Brook, 
C., Macci{\`o}, A.~V., et al.\ 2013, \mnras, 428, 129 

%NIHAO III: the constant disc gas mass conspiracy
\bibitem[Stinson et al.(2015)]{Stinson15} Stinson, G.~S., Dutton, A.~A., Wang, L., et al.\ 2015, \mnras, 454, 1105 

%Cusp-core transformations in dwarf galaxies: observational predictions
\bibitem[Teyssier et al.(2013)]{Teyssier13} Teyssier, R., Pontzen, A., Dubois, Y., \& Read, J.~I.\ 2013, \mnras, 429, 3068 

%NIHAO IV: Core creation and destruction in dark matter density profiles across cosmic time
\bibitem[Tollet et al.(2016)]{Tollet16} Tollet, E., Macci{\`o}, A.~V., Dutton, A.~A., et al.\ 2016, \mnras, 456, 3542 

%Is There Evidence for Flat Cores in the Halos of Dwarf Galaxies? The Case of NGC 3109 and NGC 6822
\bibitem[Valenzuela et al.(2007)]{Valenzuela07} Valenzuela, O., Rhee, G., Klypin, A., et al.\ 2007, \apj, 657, 773 
  
%Introducing the Illustris Project: simulating the coevolution of dark and visible matter in the Universe
\bibitem[Vogelsberger et al.(2014)]{Vogelsberger14} Vogelsberger, M., Genel, S., Springel, V., et al.\ 2014, \mnras, 444, 1518 
  
%NIHAO project - I. Reproducing the inefficiency of galaxy formation across cosmic time with a large sample of cosmological hydrodynamical simulations
\bibitem[Wang et al.(2015)]{Wang15} Wang, L., Dutton, A.~A., Stinson, G.~S., et al.\ 2015, \mnras, 454, 83 

%  Gasoline2: a modern smoothed particle hydrodynamics code
\bibitem[Wadsley et al.(2017)]{Wadsley17} Wadsley, J.~W., Keller, B.~W., \& Quinn, T.~R.\ 2017, \mnras, 471, 2357 

%Bar-driven Dark Halo Evolution: A Resolution of the Cusp-Core Controversy
\bibitem[Weinberg \& Katz(2002)]{Weinberg02} Weinberg, M.~D.,
  \& Katz, N.\ 2002, \apj, 580, 627

%Modeling the Effects of Star Formation Histories on Hα and Ultraviolet Fluxes in nearby Dwarf Galaxies
\bibitem[Weisz et al.(2012)]{Weisz12} Weisz, D.~R., Johnson, B.~D., Johnson, L.~C., et al.\ 2012, \apj, 744, 44 

%Dynamical Blueprints for Galaxies  
\bibitem[Widrow et al.(2008)]{Widrow08} Widrow, L.~M., Pym, B., \& Dubinski, J.\ 2008, \apj, 679, 1239-1259 

%The Milky Way's Circular Velocity Curve to 60 kpc and an Estimate of the Dark Matter Halo Mass from the Kinematics of ~2400 SDSS Blue Horizontal-Branch Stars  
\bibitem[Xue et al.(2008)]{Xue08} Xue, X.~X., Rix, H.~W., Zhao, G., et al.\ 2008, \apj, 684, 1143 
    
\end{thebibliography}
\end{document}